\documentclass[fleqn,10pt]{SelfArx} 

\definecolor{color1}{RGB}{0,0,90} 
\definecolor{color2}{RGB}{0,20,20} 

\usepackage{hyperref} 
\hypersetup{hidelinks,colorlinks,breaklinks=true,urlcolor=color2,citecolor=color1,linkcolor=color1,bookmarksopen=false,pdftitle={Title},pdfauthor={Author},urlcolor=blue}

\usepackage{graphicx}
\usepackage[square]{natbib}
\usepackage{amsmath}
\usepackage{multirow}
\usepackage{subfigure}
\usepackage{pdflscape}

\newcommand{\n}[1]{\mathrm{#1}}

\JournalInfo{Published in International Journal of Refrigeration, Vol. 34, 192-203, 2011} 
\Archive{\href{http://dx.doi.org/10.1016/j.ijrefrig.2010.07.004}{DOI: 10.1016/j.ijrefrig.2010.07.004}} 

\PaperTitle{The influence of the magnetic field on the performance of an active magnetic regenerator (AMR)} 

\Authors{R. Bj\o{}rk and K. Engelbrecht} 
\affiliation{\textit{Department of Energy Conversion and Storage, Technical University of Denmark - DTU, Frederiksborgvej 399, DK-4000 Roskilde, Denmark}} 
\affiliation{*\textbf{Corresponding author}: rabj@dtu.dk} 

\newcommand{\keywordname}{Keywords} 

\Abstract{The influence of the time variation of the magnetic field, termed the magnetic field profile, on the performance of a magnetocaloric refrigeration device using the active magnetic regeneration (AMR) cycle is studied for a number of process parameters for both a parallel plate and packed bed regenerator using a numerical model. The cooling curve of the AMR is shown to be almost linear far from the Curie temperature of the magnetocaloric material. It is shown that a magnetic field profile that is 10\% of the cycle time out of sync with the flow profile leads to a drop in both the maximum temperature span and the maximum cooling capacity of 20-40\% for both parallel plate and packed bed regenerators. The maximum cooling capacity is shown to depend very weakly on the ramp rate of the magnetic field. Reducing the temporal width of the high field portion of the magnetic field profile by 10\% leads to a drop in maximum temperature span and maximum cooling capacity of 5-20\%. An increase of the magnetic field from 1 T to 1.5 T increases the maximum cooling capacity by 30-50\% but the maximum temperature span by only 20-30\%. Finally, it was seen that the influence of changing the magnetic field was more or less the same for the different regenerator geometries and operating parameters studied here. This means that the design of the magnet can be done independently of the regenerator geometry.}

\begin{document}

\flushbottom 

\maketitle 


\thispagestyle{empty} 

\section{Introduction}
Magnetic refrigeration is an evolving cooling technology that has the potential for high energy efficiency using environmentally friendly refrigerants. Refrigeration is generated by utilizing the magnetocaloric effect (MCE), which is the temperature change that most magnetic materials exhibit when subjected to a changing magnetic field. This temperature change is called the adiabatic temperature change, $\Delta{}T_\mathrm{ad}$, and is a function of temperature and magnetic field. The temperature change is greatest near the Curie temperature, $T_\mathrm{c}$, which varies with the magnetocaloric material \citep{Pecharsky_2006}. Because the MCE in the best magnetocaloric materials currently available exhibit a temperature change of no more than 4 K in a magnetic field of 1 T \citep{Dankov_1998}, a magnetic refrigeration device must utilize a regenerative process to produce a large enough temperature span to be useful for refrigeration purposes. The most utilized process for this is called active magnetic regeneration (AMR) \citep{Barclay_1982}.

A great number of magnetic refrigeration test devices have been built and examined in some detail, with focus on the produced temperature span and cooling power of the devices \citep{Barclay_1988, Yu_2003,Gschneidner_2008}. Detailed and extensive investigations of the AMR process using numerical modeling have previously been published \citep{Hu_1995,Engelbrecht_2005a,Engelbrecht_2005b,Allab_2005,Siddikov_2005,Shir_2005b,Petersen_2008a,Nielsen_2009a}, but so far little focus has been put into investigating how the properties and time variation of the magnetic field influence the theoretical performance of the AMR cycle. Here, a generic magnetic field that varies as a function of time during the AMR cycle is used to investigate the influence of the magnetic field on the performance of the AMR process. This time varying profile is called the magnetic field profile.

\subsection{The AMR process}
In the AMR process a heat transfer fluid and a magnetocaloric material (MCM), acting as a regenerator, are used to build up a temperature gradient that can be much larger than the adiabatic temperature change produced by the magnetocaloric material. The regenerator consists of a porous matrix of a solid magnetocaloric material through which a non-magnetic fluid can flow. This fluid transfers heat (positive or negative) to the solid material and through a movement of the fluid, by a piston or a pump, it is moved to heat exchangers in a cooled space or in contact with the environment. Most AMR devices either have a packed bed regenerator where the MCM is typically packed spheres \citep{Okamura_2005,Tura_2009} or a parallel plate regenerator \citep{Zimm_2007, Bahl_2008}. For a review of different magnetic refrigeration devices please see \citet{Gschneidner_2008}.

An AMR cycle proceeds in four steps. First the regenerator is magnetized. This raises the temperature of the solid due to the magnetocaloric effect.  The temperature rise is a function of magnetic field but also of temperature, and thus of position in the regenerator. At the same time heat is transferred from the MCM to the heat transfer fluid in a time span, $\tau_{1}$. The heat transfer fluid is then displaced towards the hot heat exchanger where the heat is released to the surroundings over a time span, $\tau_{2}$. Next, the magnetic field in the regenerator is removed. This lowers the temperature of the MCM by the adiabatic temperature change so that the MCM is now colder than the entrained heat transfer fluid. Thus heat is transferred from the heat transfer fluid to the MCM, cooling the heat transfer fluid, through a time span, $\tau_{3}$. Then the heat transfer fluid is displaced towards the cold heat exchanger, where heat can be absorbed from a heat load through a time span, $\tau_{4}$. A total cycle lasts a time span $\tau$, equal to $\tau_{1}+\tau_{2}+\tau_{3}+\tau_{4}$. The process then starts over again. Using this regenerative process a temperature span between the hot and cold end that is greater than the adiabatic temperature change can be achieved.

The performance of an AMR device depends on the process parameters specific to each AMR system. These are the shape and packing of the magnetocaloric material, the temperature of the surroundings and the properties of the MCM used, as well as the properties of the heat transfer fluid, flow system etc. The only parameter that is common to all AMR systems is the magnetic field. The magnet might be shaped differently in different AMR system, but the magnetic field generated by the magnet has the same effect on all AMR systems, namely that it drives the magnetocaloric effect that is the heart of the AMR. Thus the magnetic field sets a fundamental limit on the temperature span and cooling power that an AMR system can produce. Therefore it is interesting to investigate the influence of the magnetic field on the AMR performance.

\begin{figure}[!t]
  \centering
  \includegraphics[width=1.0\columnwidth]{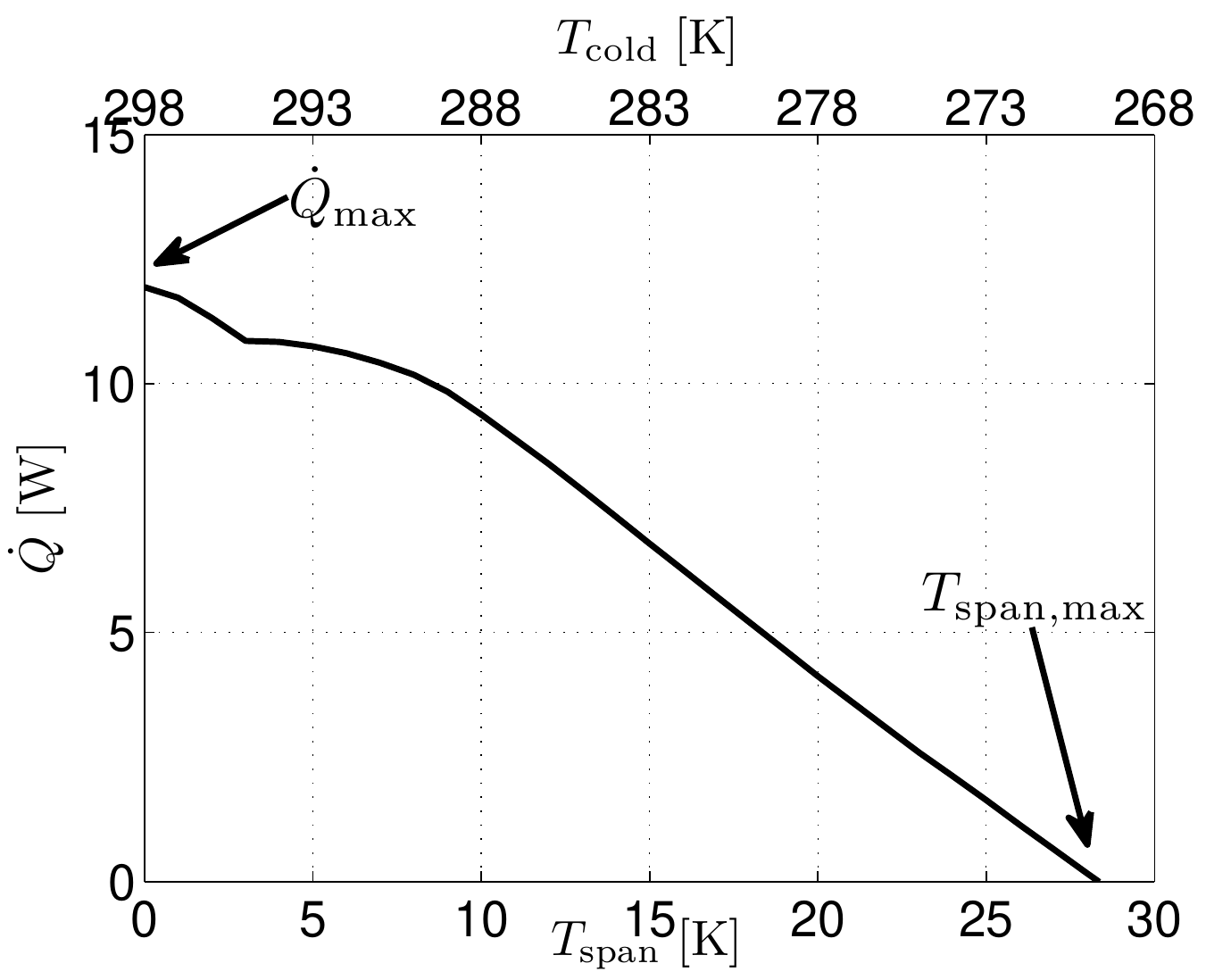}
  \caption{An example of a cooling curve showing $\dot{Q}$ as a function of $T_\n{span}$. The maximum temperature span, $T_\n{span,max}$, and the maximum refrigeration capacity, $\dot{Q}_\n{max}$, have been indicated. The temperature of the hot end of the regenerator was $T_\n{hot} = 298$ K and the Curie temperature of the MCM of $T_\n{C} = 293.6$ K.}
\label{Fig_Cooling_curve}
\end{figure}

The performance of an AMR is summed up in the cooling curve of the AMR. This curve shows the cooling capacity, $\dot{Q}$, as a function of temperature span, $T_\n{span}$, of the device, for a given set of process parameters. The temperature span is the difference between the temperature of the hot and the cold end, $T_\mathrm{hot}$ and $T_\mathrm{cold}$, respectively. An example of a cooling curve is shown in Fig. \ref{Fig_Cooling_curve}. Examination of the figure shows that the cooling power produced by the AMR is highly dependent on temperature span. The most often-cited information that can be learned from a cooling curve are the maximum or no load temperature span, $T_\n{span,max}$, and the maximum refrigeration capacity, $\dot{Q}_\n{max}$. At $T_\n{span,max} \Leftrightarrow{} \dot{Q} = 0$ W, while at $\dot{Q}_\n{max} \Leftrightarrow{} T_\n{span} = 0$ K, where in the latter case all the cooling power generated by the device is used to move heat from the cold to the hot end. These two parameters are used to characterize the efficiency of an AMR throughout this paper. The shape of the cooling curve in between these two points is of course also of interest and is also investigated here.

The effect of the magnetic field on the performance of the AMR cycle is controlled by three parameters. The first of these parameters is the synchronization of the magnetic field with the AMR cycle, i.e. when in the AMR cycle the magnetocaloric material is subjected to the magnetic field. The second parameter is the ramp rate of the magnetic field, i.e. how quickly does the magnetic field change from its minimum to its maximum value and vice versa. And finally the last parameter is the maximum value of the magnetic field. A spatial variation of the magnetic field across the AMR is not considered here.

The influence of each of these parameters on the performance of the AMR are studied using a numerical model for a number of different set of AMR process parameters. Ideally the work presented here should be supported by experimental data, but conducting AMR experiments with changing magnetic fields are notoriously cumbersome, as most magnetic refrigerators use permanent magnets to generate the magnetic field, and for these the generated magnetic flux density can rarely be changed. Experiments would be possible if an electromagnet or an adjustable permanent magnet assembly was used, as in the AMR devices by \citet{Tura_2007,Bahl_2008}.

\section{The numerical model}
A one dimensional numerical model capable of modeling both packed bed and parallel plate regenerators is used to model the AMR process \citep{Engelbrecht_2006}. This model is publicly available. For the packed bed regenerator the model has previously been compared with experimental data \citep{Engelbrecht_2008}. For the parallel plate regenerator case the model has been compared with a more detailed two dimensional model \citep{Petersen_2008b}, where the latter has been compared with experimental data \citep{Bahl_2008}. In the numerical model, the temperature span is an input parameter and the refrigeration capacity is calculated for the specified process parameters.

The one dimensional model assumes that the fluid and solid temperature profiles are functions only of the flow direction. The cooling capacity of the AMR is determined by solving the coupled one-dimensional partial differential equations in space and time describing the temperature in the regenerator and in the fluid. Different regenerator parameters such as the position dependent Nusselt number, which determines the heat transfer between the regenerator and the fluid, and the friction factor are determined using established correlations. The model assumes that the edges of both the fluid and the solid are adiabatic except during the blow periods where the fluid enters the regenerator with the prescribed temperature of either the hot or the cold reservoir. The model starts from an initial temperature distribution and takes time steps until a cyclical steady state has been achieved. This state is reached when the dimensionless value of the absolute change in energy of the regenerator from cycle to cycle is less than a specified tolerance. The governing equations for the model are given in \citet{Engelbrecht_2006,Petersen_2008b}.

For the parallel plate regenerator model the comparison with the two dimensional model lead to the definition of a ``1D correctness'' parameter, $\Gamma$, which is defined as
\begin{eqnarray}\label{Eq.1D_correctness_parameter}
\Gamma{}=\frac{\pi^2k_\mathrm{fluid}}{h^2_\mathrm{fluid}\rho{}_\mathrm{fluid}c_\mathrm{p, fluid}}\tau_1\gg1
\end{eqnarray}
where the subscript ``fluid'' denotes a property of the heat transfer fluid, and where $k$ is the thermal conductivity, $\rho{}$ is the mass density, $c_\mathrm{p}$ is the specific heat capacity and $h_\mathrm{fluid}$ is the height of the fluid channel.  A value much greater than one for $\Gamma$ corresponds to an operating condition in which the one dimensional model produces comparable results to the two dimensional model.


\subsection{The magnetic field profile}
\begin{figure}[!t]
  \centering
  \includegraphics[width=1.0\columnwidth]{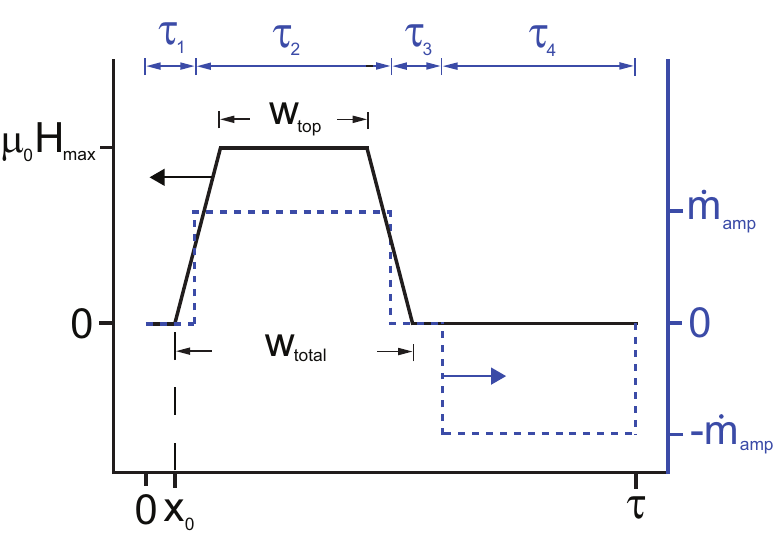}
\caption{The generic magnetic field profile (full line) and the fluid flow cycle (dotted line) of the AMR cycle. The parameters of the magnetic field profile are indicated, as well as the AMR cycle time parameters $\tau_{1-4}$.}
\label{Fig_Ramping_curve}
\end{figure}

To separately study the effects of each of the properties of the magnetic field, a generic magnetic field profile is used. This profile is shown in Fig. \ref{Fig_Ramping_curve} along with the fluid flow profile.

The generic magnetic field profile can be characterized by four parameters. The first parameter is the synchronization, denoted by $x_0$, which indicates the time in fractions of $\tau$ where the magnetic field profile begins to increase, relative to the AMR fluid flow cycle. The second parameter is the maximum value of the magnetic field, $\mu_0{}H_\n{max}$, which is in general varied between 0.5 and 1.5 T, as this is the range obtainable with permanent magnets \citep{Bjoerk_2010b}. The final two parameters are the top width, $w_\n{top}$, and the total width, $w_\n{total}$, of the temporal magnetic field profile, which are both defined in terms of $\tau$. These also define the ramp rate.

\section{AMR process parameters}
The performance of the AMR as a function of magnetic field is studied for a number of process parameters. These must be chosen realistically if the results of the numerical model are to be relevant for magnetic refrigeration test devices. In all models and for all process parameters, a symmetric, or balanced, AMR flow cycle is used. The values of the fluid flow cycle parameters are always $\tau_1=\tau_3=0.1$ and $\tau_2=\tau_4=0.4$ in fractions of the total cycle time, $\tau$. The length of the modeled regenerator is 50 mm. The heat transfer fluid is water with constant properties as given in \citet{Petersen_2008b}. The MCM is taken to be gadolinium, modeled using the mean field model \citep{Morrish_1965} with a Curie temperature of $T_\n{c} = 293.6$ K and properties as given in \citet{Petersen_2008a}. Although the mean field model does not exactly reproduce experimental data \citep{Dankov_1998, Liu_2007}, it is often used as the benchmark model for magnetocaloric AMR models \citep{Petersen_2008a,Nielsen_2010} because it produces thermodynamically consistent data with smooth derivatives  and different numerical models can be more easily compared if the same data set has been used as input. The temperature of the hot end of the AMR is kept fixed at $T_\n{hot}=298$ K.

For the parallel plate regenerator three process parameters must be specified. These are the height of the fluid channel, $h_\mathrm{fluid}$, the height of the plate, $h_\mathrm{plate}$, and the mass flow rate, $\dot{m}_\mathrm{amp}$. Here 54 different sets of parameters are considered. These are listed in Table \ref{Table.Parallel_plate_parameter_sets}, and have been chosen so that they span realistic values of the different parameters and yet produce similar results to the two dimensional model mentioned previously. The mass flow rate has been chosen so that it is $7.27*10^{-3}$ kg s$^{-1}$ for a 1 mm plate at $\tau = 6$ s \citep{Petersen_2008b}. With the chosen values for the height of the plate and fluid channel the porosity is 50\%, 66\% and 75\% respectively.

\begin{table*}[!t]
\caption{Parallel plates regenerator parameters. As the parameters are varied individually, the table should not be read as rows but rather as what values the different parameters can assume. In total there are 54 sets of parameters, but 18 sets for $\tau = 0.5$ s are disregarded as the results would differ from a two dimensional AMR model.}\label{Table.Parallel_plate_parameter_sets}
\begin{center}
\begin{tabular}{c|c|c|c}
$h_\mathrm{fluid}$ [m] & $h_\mathrm{plate}$ [m] & $\dot{m}_\mathrm{amp}$ [kg s$^{-1}$] & $\tau$ [s]\\ \hline
0.00010  & $1*h_\mathrm{fluid}$ & $0.5*7.27*6/\tau{}*h_\mathrm{plate}$  & 0.5\\
0.00025  & $2*h_\mathrm{fluid}$ & $1*7.27*6/\tau{}*h_\mathrm{plate}$ & 6 \\
0.00050  & $3*h_\mathrm{fluid}$ & $2*7.27*6/\tau{}*h_\mathrm{plate}$
\end{tabular}
\end{center}
\end{table*}

For the cycle time of $\tau=0.5$ s the result of the one dimensional model might deviate from a more detailed two dimensional model, as per the $\Gamma$ parameter defined in Eq. (\ref{Eq.1D_correctness_parameter}). If $\Gamma < 3$ the set of process parameters are not considered further. For the $\tau=0.5$ s parallel plate case these are the parameter sets where $h_\n{fluid} > 0.00010$ m. Thus a total of 18 sets of parameters are disregarded for the case of $\tau=0.5$ s. For the case of $\tau = 6$ s, the lowest value of $\Gamma$ occurs for $h_\mathrm{fluid} = 0.00050$ m, where $\Gamma = 3.38$, thus all sets of parameters are within the defined requirement for $\Gamma$.

\begin{figure}[!t]
  \centering
  \includegraphics[width=1.0\columnwidth]{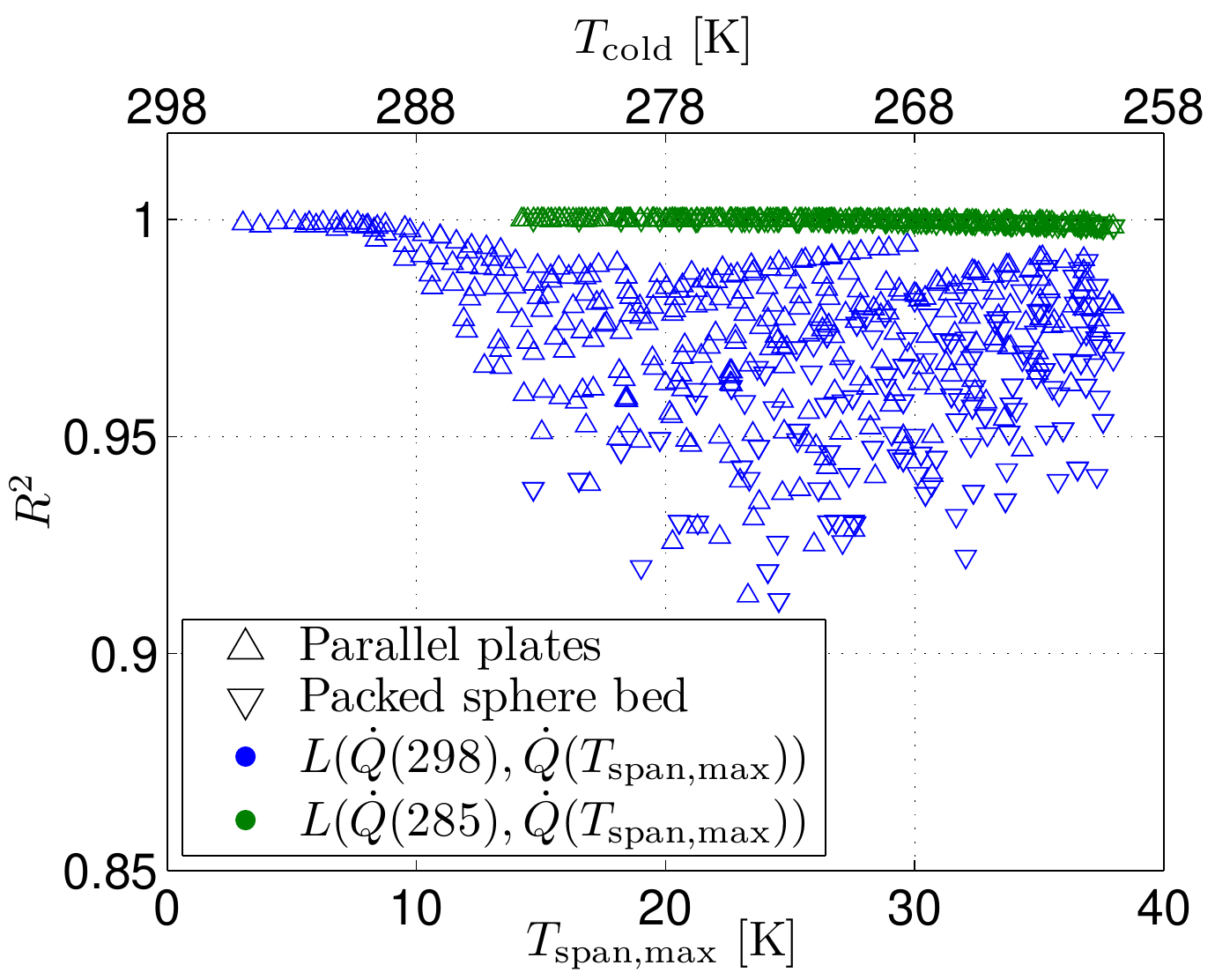}
\caption{The goodness of fit parameter $R^2$ as a function of $T_\n{span,max}$ for parallel plates and packed bed for a linear fit to the full cooling curve and the cooling curve from 285 K to $T_\n{span,max}$.}
\label{Fig_R_square_factor}
\end{figure}

A spherical particle packed bed regenerator has also been considered. Here the process parameters are the height of the regenerator, $h$, the particle size, $d_\n{p}$, the mass flow rate, $\dot{m}_\mathrm{amp}$, and the porosity. For a randomly packed sphere bed regenerator used in magnetic refrigeration the latter is generally near 0.36 \citep{Okamura_2005,Jacobs_2009, Tura_2009} and therefore this parameter is fixed. The height of the regenerator, $h$, is chosen to be identical to three of the values from the parallel plate case, resulting in an equal regenerator volume and equal utilization for these cases. The utilization is given as $\Phi = \dot{m}_\n{fluid}c_\n{p,fluid}P/m_\n{solid}c_\n{p,solid}$ where $P$ is the time period for either the hot or cold blow. All nine different heights from the parallel plate cases cannot be tried as this would result in too many parameter sets. As the height of the regenerator is increased the utilization will drop as the mass of magnetocaloric material is increased but the mass flow rate is kept constant. The particle size, $d_\n{p}$, is varied within reasonable values \citep{Okamura_2005,Engelbrecht_2007c,Tura_2009}. The value of $\dot{m}_\mathrm{amp}$ is calculated to give the same value as the parallel plate cases with the same regenerator geometry. Finally $\tau$ assumes the values of 0.5 and 6 s. The parameter sets are listed in Table \ref{Table.Packed_bed_parameter_sets}.

\begin{table*}[!t]
\caption{Packed bed regenerator parameters. Similarly to the parallel plate parameter table, Table \ref{Table.Parallel_plate_parameter_sets}, the rows in this table are not to be understood as parameter sets, except for the $\dot{m}_\mathrm{amp}$ column. Here, e.g., for the case of $h=0.0002$ m the value of $\dot{m}_\mathrm{amp}=0.0001*[0.5\;\; 1\;\; 2]*7.27*6/\tau$ kg s$^{-1}$ and not any other values. Similarly for $h=0.00075$ m the value is $\dot{m}_\mathrm{amp}=0.0005*[0.5\;\; 1\;\; 2]*7.27*6/\tau$ kg s$^{-1}$ and not any other values and so on. This ensures that $\dot{m}_\mathrm{amp}$ is equal for the parallel plate and packed bed cases with the same regenerator volume. There are a total of 54 parameter sets.}\label{Table.Packed_bed_parameter_sets}
\begin{center}
\begin{tabular}{c|c|c|c}
$h$ [m] & $d_\n{p}$ [m] & $\dot{m}_\mathrm{amp}$ [kg s$^{-1}$] & $\tau$ [s]\\ \hline
0.00020  & 0.00010 & $0.0001*[0.5\;\; 1\;\; 2]*7.27*6/\tau{}$ & 0.5 \\
0.00075  & 0.00025 & $0.0005*[0.5\;\; 1\;\; 2]*7.27*6/\tau{}$ & 6  \\
0.00150  & 0.00050 & $0.0010*[0.5\;\; 1\;\; 2]*7.27*6/\tau{}$
\end{tabular}
\end{center}
\end{table*}

\section{The cooling curve}
As previously mentioned the performance of an AMR, for a given set of process parameters, is summed up in the cooling curve which shows the cooling capacity, $\dot{Q}$, as a function of temperature span, $T_\n{span}$, of the device. The cooling curve is often assumed to be linear which would allow for interpolation to find $T_\n{span,max}$. This is useful because the model used here calculates $\dot{Q}$ for a chosen $T_\n{span}$. Here we have investigated the linearity by calculating the $R^{2}$ parameter of a linear least-squares fit to the cooling curve for a symmetric magnetic field profile, i.e. $w_\n{top} = 0.45$, $w_\n{total}=0.55$ and $x_0=0$, for all parameters sets, and $T_\n{cold}=260$ K to 298 K in steps of 1 K and at $\mu_0H_\n{max}=$ 0.5 T to 1.5 T in steps of 0.1 T for both the parallel plates and packed bed regenerator. The parameter $R^{2}$ is a goodness-of-fit statistic that measures of how well the fit approximates the data points, with an $R^{2}$ value of 1.0 indicating a perfect fit. The parameter is defined as
\begin{eqnarray}
R^2\equiv{}1-\frac{\sum_{i}(y_i-f_i)^2}{\sum_{i}(y_i-\bar{y})^2},
\end{eqnarray}
where $y_i$ are the values of the data set, $f_i$ are the interpolated values and $\bar{y}$ is the mean of the data.

The shape of a typical cooling curve displays a flattening around the Curie temperature. Here the temperature of the hot end, $T_\n{hot} = 298$ K, is larger than the Curie temperature of the MCM, $T_\n{c} = 293.6$ K, and thus this flattening will be present, as can also be seen in the cooling curve shown in Fig. \ref{Fig_Cooling_curve}.

The $R^{2}$ parameter is shown in Fig. \ref{Fig_R_square_factor} as a function of $T_\n{span,max}$ for two different linear functions. For this figure $T_\n{span,max}$ has been determined fairly accurately because the temperature has been varied in steps of 1 K. The functions are linear fits of the cooling curve from 298 K, i.e. at $\dot{Q}_\n{max}$, to $T_\n{span,max}$ and a linear fit from 285 K to $T_\n{span,max}$. The first fit is denoted $L(\dot{Q}(298),\dot{Q}(T_\mathrm{span,max}))$ and the latter \\ $L(\dot{Q}(285),\dot{Q}(T_\mathrm{span,max}))$ in Fig. \ref{Fig_R_square_factor}. As can clearly be seen from the figure the cooling curve is not linear from $\dot{Q}_\n{max}$ to $T_\n{span,max}$. However, if only the part after the flattening, i.e. $L(\dot{Q}(285),\dot{Q}(T_\mathrm{span,max}))$, is fitted the $R^{2}$ parameter is in general higher than 0.998, and thus this part of the cooling curve is extremely close to being linear for both parallel plate and packed bed regenerators.

\begin{figure}[!t]
  \centering
  \includegraphics[width=1.0\columnwidth]{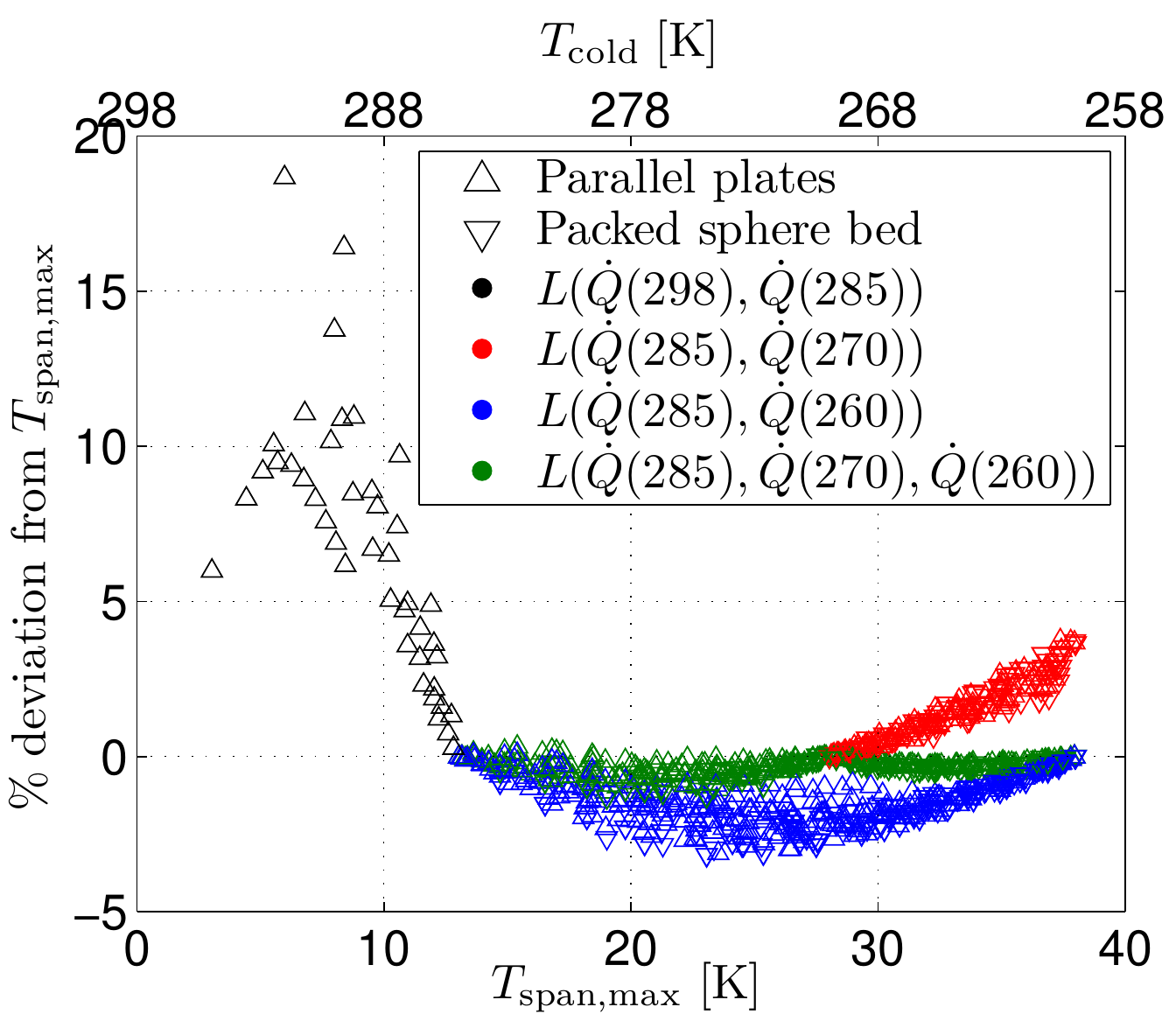}
\caption{The deviation of the estimated $T_\n{span,max}$, based on different linear fits, from the true $T_\n{span,max}$. From $T_\n{cold}=$ 285 K to 270 K the two linear functions $L(\dot{Q}(285),\dot{Q}(270))$ and $L(\dot{Q}(285),\dot{Q}(270),\dot{Q}(260))$ are identical, and thus only $L(\dot{Q}(285),\dot{Q}(270),\dot{Q}(260))$ is shown.}
\label{Fig_Linearity_interpolation}
\end{figure}

Having shown that the part of the cooling curve away from $T_\n{c}$ can be fitted by a linear function, we wish to examine if the $T_\n{span,max}$ parameter can be estimated by only calculating the refrigeration capacity at a few selected temperature spans. In Fig. \ref{Fig_Linearity_interpolation} four different linear functions have been used to calculate $T_\n{span,max}$, i.e. where $\dot{Q} = 0$. The first is a linear function between the two points $\dot{Q}(T_\n{cold} = 298)$ and $\dot{Q}(T_\n{cold} = 285)$, denoted $L(\dot{Q}(298),\dot{Q}(285))$. Using the same notation the three remaining functions are $L(\dot{Q}(285),\dot{Q}(260))$, $L(\dot{Q}(285),\dot{Q}(270))$ and $L(\dot{Q}(285),\dot{Q}(270),\dot{Q}(260))$, where the latter is a two-section linear function, which consists of a linear function from $\dot{Q}(285)$ to $\dot{Q}(270)$ and second from $\dot{Q}(270)$ to $\dot{Q}(260)$. Thus for the two-section linear function if $\dot{Q}(270)$ is negative $T_\n{span,max}$, is interpolated between $\dot{Q}(285)$ to $\dot{Q}(270)$ while the $\dot{Q}(270)$ is positive the linear function $\dot{Q}(270)$ to $\dot{Q}(260)$ is used. A negative value of $\dot{Q}$ simply means that the regenerator is not able to cool the heat load sufficiently. The $T_\n{span,max}$ parameter has been interpolated based on these linear functions and the deviation from the true $T_\n{span,max}$, found by linear interpolation of the data set where $T_\n{cold}$ was varied in steps of 1 K, is shown in Fig. \ref{Fig_Linearity_interpolation}. If the value of $T_\n{span,max}$ is higher than $38$ K i.e. so that $\dot{Q}(260)$ is positive, than the value is disregarded. This is the case for the remainder of this article.

From the figure it can be seen that all the linear functions provide an accurate estimate of $T_\n{span,max}$. However, the two-section linear function $L(\dot{Q}(285),\dot{Q}(270),\dot{Q}(260))$ provides the most precise estimate of $T_\n{span,max}$, i.e. one that is accurate to within 1.5\% of the true value in the $T_\n{span,max}$ interval between 285 K and 260 K. In the interval between 298 K and 285 K the estimate of $T_\n{span,max}$ is not very accurate, but as there are few models that have this small a temperature span the linear interpolation will still be used. In the remainder of this work the refrigeration capacity will be calculated at $T_\n{cold}=$ 298, 285, 270 and 260 K, and $T_\n{span,max}$ will be estimated based on these values using linear interpolation, unless otherwise stated.

\begin{figure*}[!t]
  \begin{center}
\subfigure[Parallel plates.]{\includegraphics[width=0.47\textwidth]{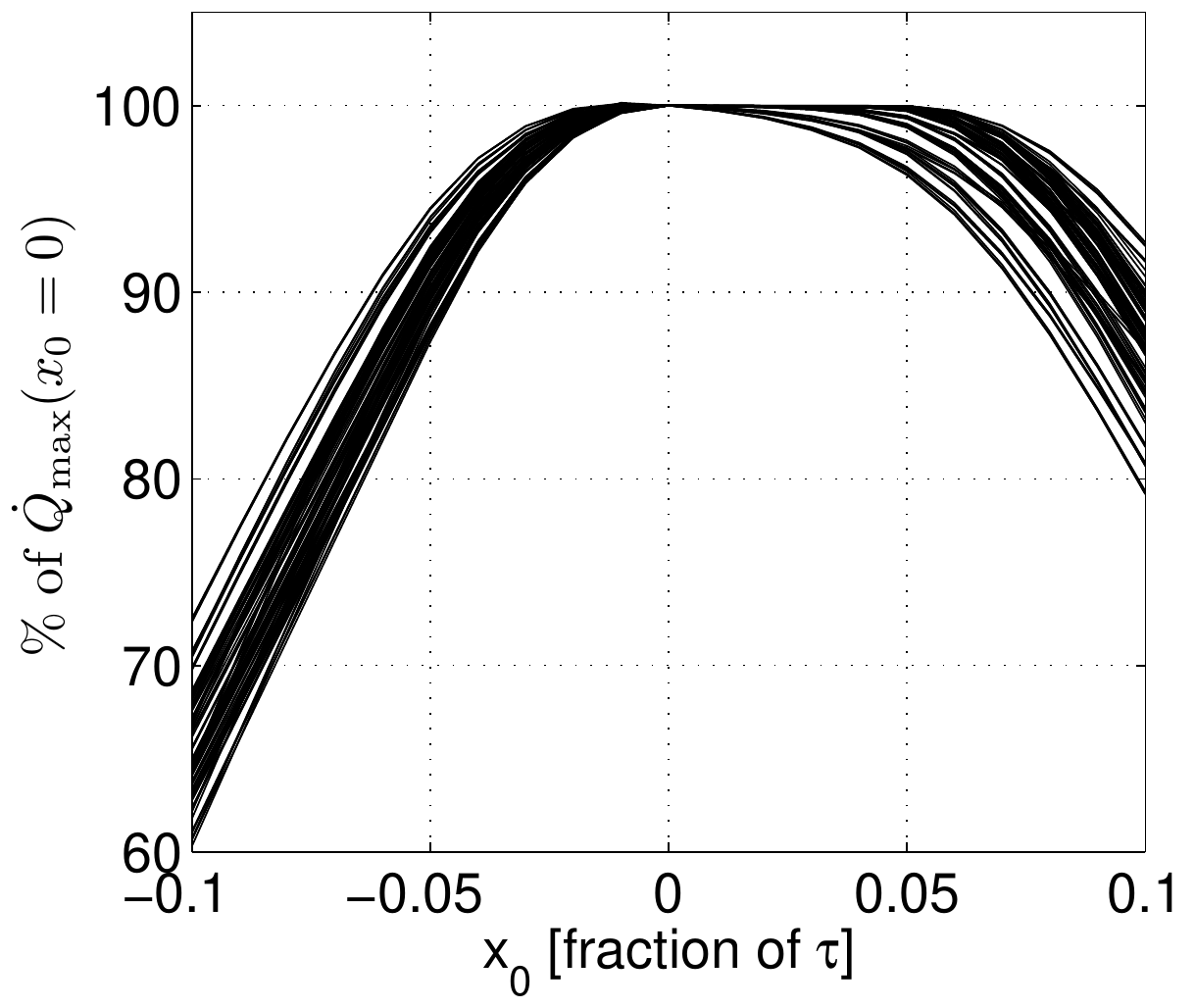}}\hspace{0.4cm}
\subfigure[Packed bed.]{\includegraphics[width=0.47\textwidth]{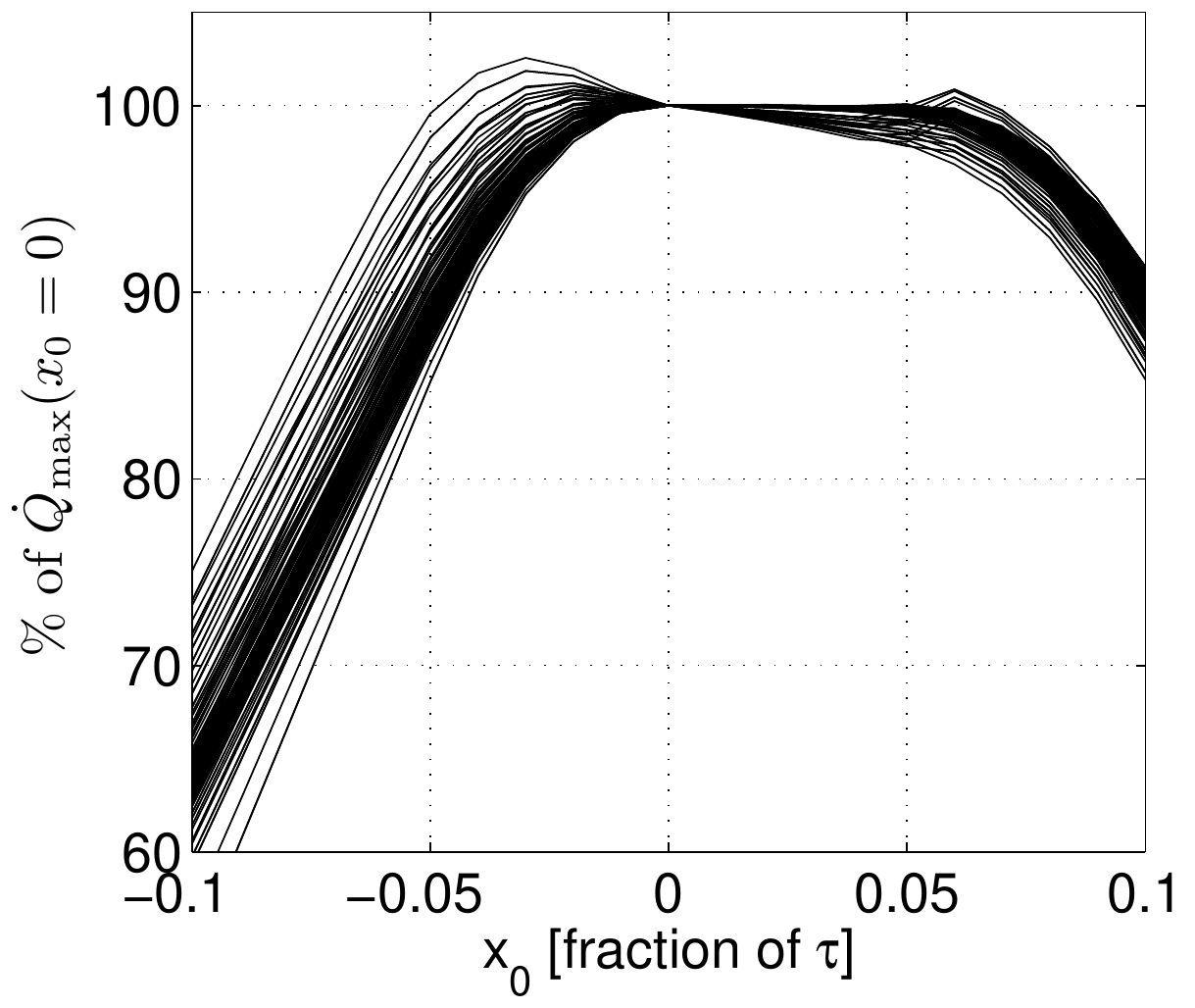}}
\end{center}
\caption{The maximum cooling capacity, $\dot{Q}_\n{max}$, as a function of the synchronization parameter, $x_0$, for all parameter sets.}
\label{Fig.Shift_Q_dot}
\end{figure*}

\begin{figure*}[!t]
\begin{center}
\subfigure[Parallel plates.]{\includegraphics[width=0.47\textwidth]{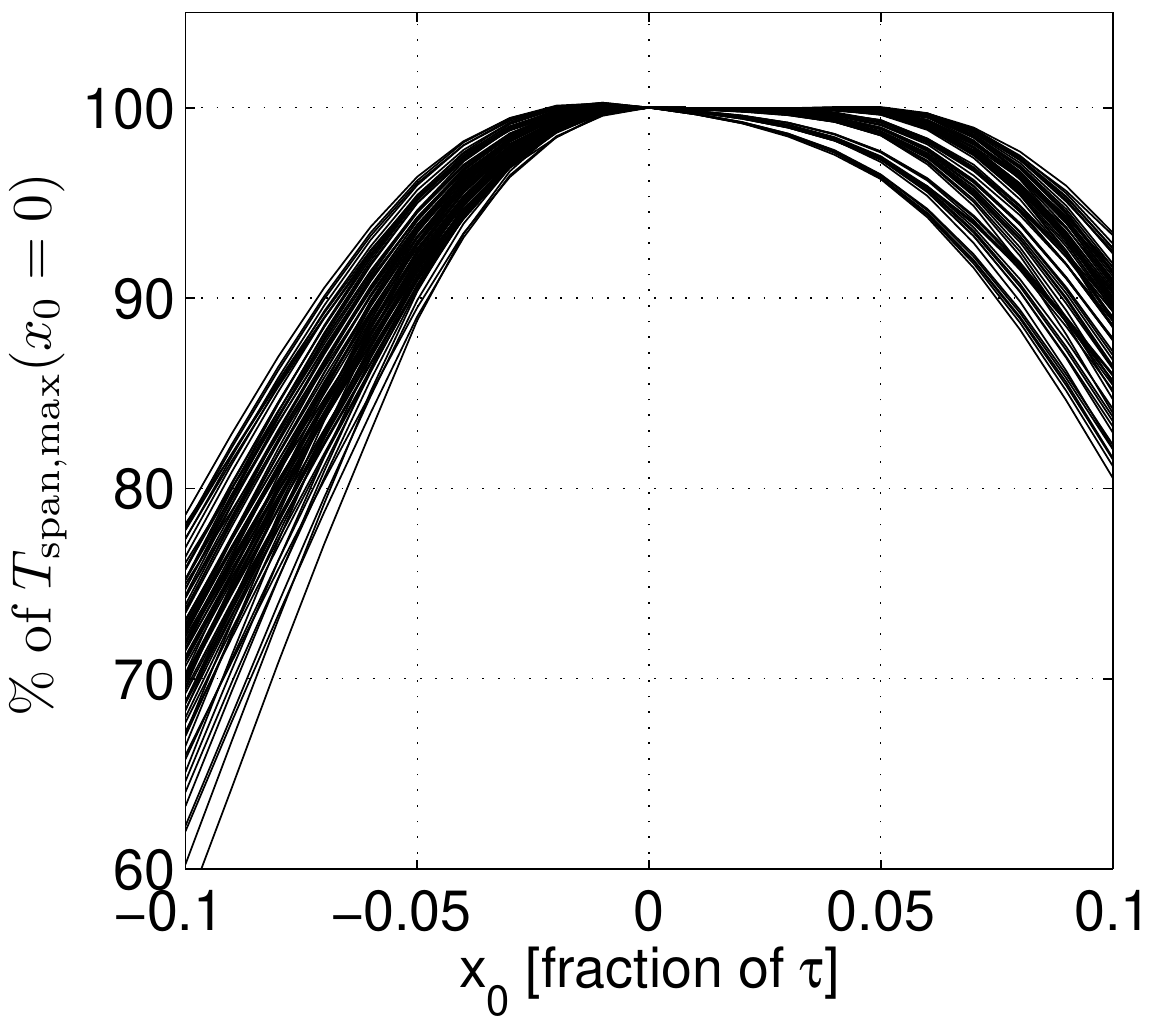}}\hspace{0.4cm}
\subfigure[Packed bed.]{\includegraphics[width=0.47\textwidth]{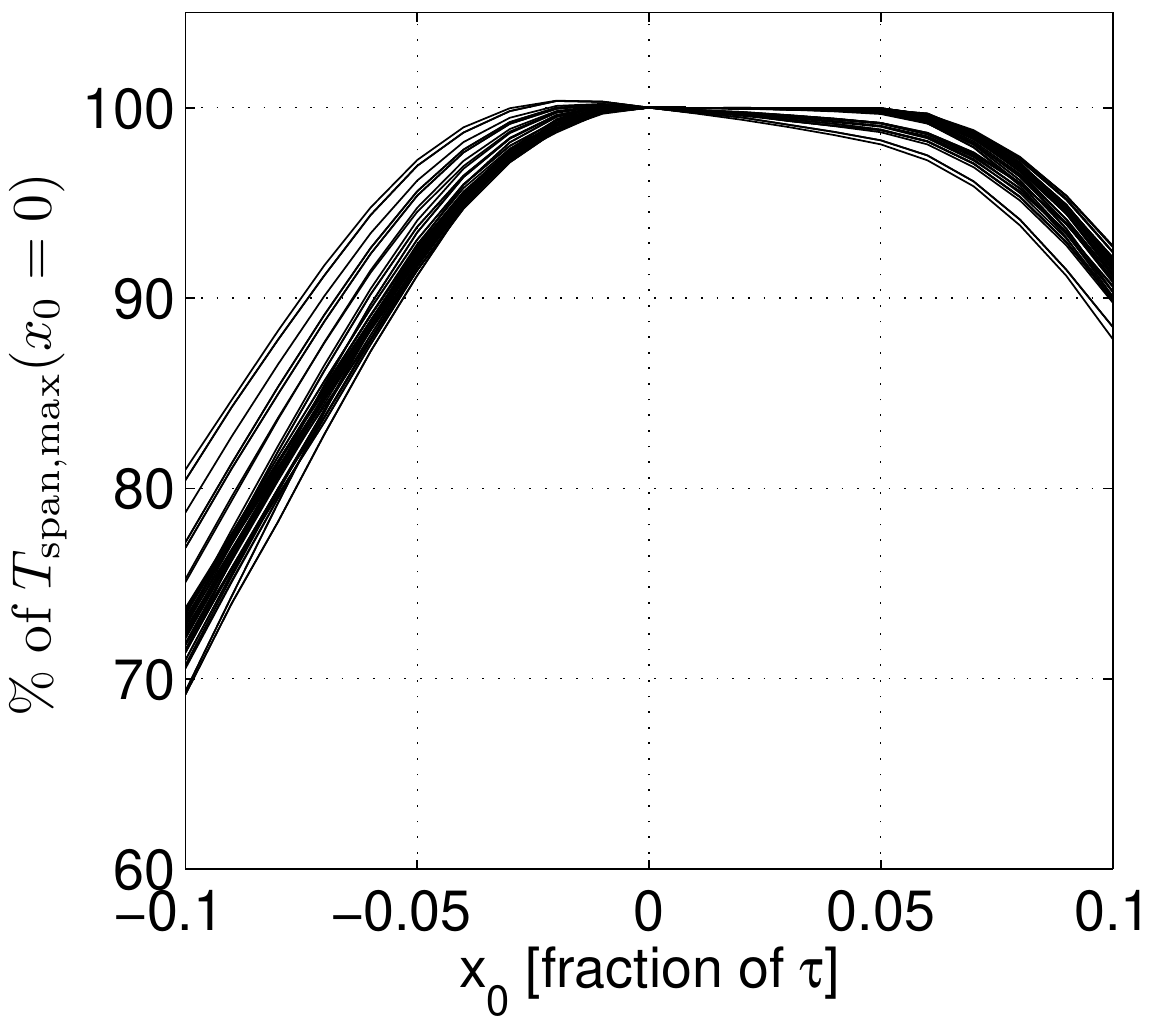}}
\end{center}
\caption{The no load temperature span, $T_\n{span,max}$, as a function of the synchronization parameter, $x_0$, for all parameter sets.}
\label{Fig.Shift_T_span}
\end{figure*}

\section{Synchronization of the magnetic field}
It is important to investigate the performance of the AMR cycle as a function of when in the AMR cycle the regenerator is subjected to the magnetic field. This is termed ``synchronization'' as it describes how synchronized the magnetic field profile is with the AMR fluid flow profile and it is characterized by the synchronization parameter, $x_0$, as shown previously in Fig. \ref{Fig_Ramping_curve}. In this article, a synchronized cycle occurs when the magnetic field profile begins to ramp up when $\tau_1$ begins and begin to ramp down when $\tau_2$ ends. A negative value of $x_0$ means that the magnetic field profile has been moved so that it begins earlier in the cycle compared to the synchronized case.

The effect of the synchronization on $\dot{Q}_\n{max}$ and $T_\n{span,max}$ has been investigated for all parameter sets for both the parallel plate and packed bed cases. A value of $\mu_0{}H_\n{max} = 0.5$, 1 and 1.5 T and a magnetic field profile with $w_\n{top} = 0.45$ and $w_\n{total}=0.55$ were used in the synchronization study. The synchronization parameter, $x_0$, was varied from -0.1 to 0.1 in steps of 0.01, where 0 is the point at which the magnetic field profile is synchronized with respect to the AMR cycle for the values of $w_\mathrm{top}$ and $w_\mathrm{total}$ used here.

The results are shown in Figs. \ref{Fig.Shift_Q_dot} and \ref{Fig.Shift_T_span} which show $\dot{Q}_\n{max}$ and $T_\n{span,max}$ as a function of the synchronization parameter, $x_0$, for all process parameters and magnetic fields. As can be seen from the figures both $\dot{Q}_\n{max}$ and $T_\n{span,max}$ show a broad plateau around $x_0 = 0$. The behavior of $\dot{Q}_\n{max}$ and $T_\n{span,max}$ appears to be much the same across the different sets of AMR process parameters and different magnetic fields that were investigated. The parallel plates show a slightly larger drop in $T_\n{span,max}$ than the packed bed regenerator does. For all parameter sets a drop in performance between 20 and 40 \% is seen if the magnetic field is ramped too early. A smaller performance drop is seen if the field is ramped too late.  Shifting the magnetic field to earlier in the cycle can also provide a minimal increase in performance.  Figures \ref{Fig.Shift_Q_dot} and \ref{Fig.Shift_T_span}  show that synchronization of the magnetic field and fluid flow is important; however, it has a small effect when the synchronization is within 5\% of the cycle time.

\section{Ramp rate of the magnetic field}
It is also important to investigate the influence of the ramp rate, i.e. the speed of the increase from $\mu_0H = 0$ T to $\mu_0H_\n{max}$, on the performance of the AMR cycle. The ramp rate can be controlled by varying the $w_\n{top}$ and $w_\n{total}$ parameters as the ramp rate is given by $\mathrm{Ramp Rate} = (w_\n{total}-w_\n{top})/2$ in fractions of $\tau$.

In the simulations $w_\n{top}$ is kept constant at $0.45$ and $\mu_0{}H_\n{max} = 1$ T. The total width, $w_\n{total}$, was varied from 0.46 to 0.65 in steps of 0.01, which means that the ramp rate was varied from 0.005 to 0.1 in fractions of $\tau$. The synchronization parameter, $x_0$, was chosen such that the magnetic field profile always begins to ramp down at $t=\tau_1+\tau_2$. In practice this means that the synchronization parameter was given as $x_0=(0.55-w_\n{total})/2$. It is ensured that the ramp of the magnetic field is always sufficiently numerically resolved in time.

\begin{figure}[!t]
\begin{center}
\subfigure[Parallel plates.]{\includegraphics[width=0.47\textwidth]{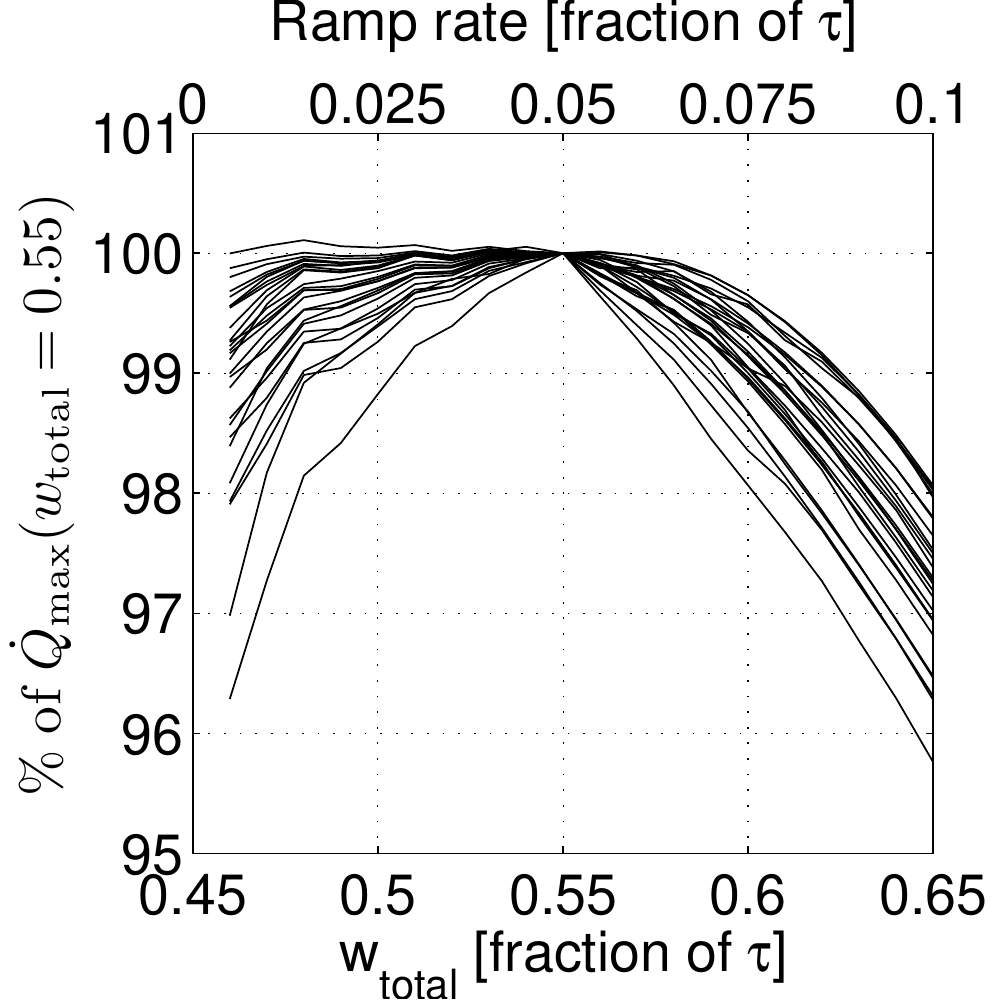}}\hspace{0.4cm}
\subfigure[Packed bed.]{\includegraphics[width=0.47\textwidth]{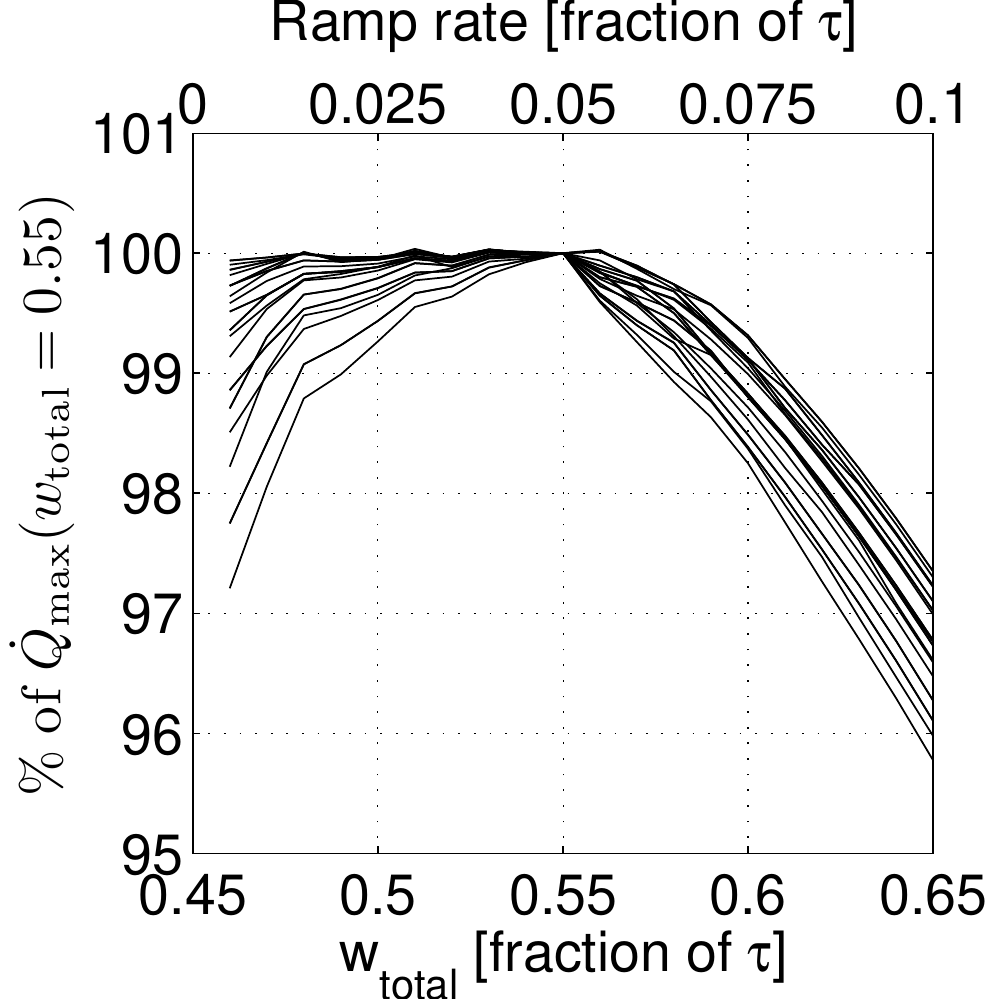}}
\end{center}
\caption{The maximum cooling capacity, $\dot{Q}_\n{max}$, as a function of $w_\n{total}$ for the parallel plate and packed bed cases. The top x-axis show the corresponding ramp rate. The lack of smoothness of the curves is due to the very small change of the refrigeration capacity with ramp rate, which is hard to resolve numerically.}
\label{Fig.Ramping_experiment_Q_dot}
\end{figure}

The behavior of the AMR system, which was only examined at the maximum refrigeration capacity, is shown in Fig. \ref{Fig.Ramping_experiment_Q_dot} which shows $\dot{Q}_\n{max}$ as a function of $w_\n{total}$ and the ramp rate for all process parameters. The maximum temperature span was not found due to the extensive computation time caused by the high numerical resolution required to resolve the field profiles with fast ramp rates. It is seen that the decrease in performance is only a few percent if a slow ramp rate is used. For a fast ramp rate no gain in performance is seen, but in a few cases a drop in performance is observed. This drop in performance can occur because the magnetic field profile with a slow ramp rate is wider, which can improve performance for some sets of process parameters. This will be considered in more detail shortly.

\begin{figure*}[!t]
\begin{center}
\subfigure[Parallel plates.]{\includegraphics[width=0.47\textwidth]{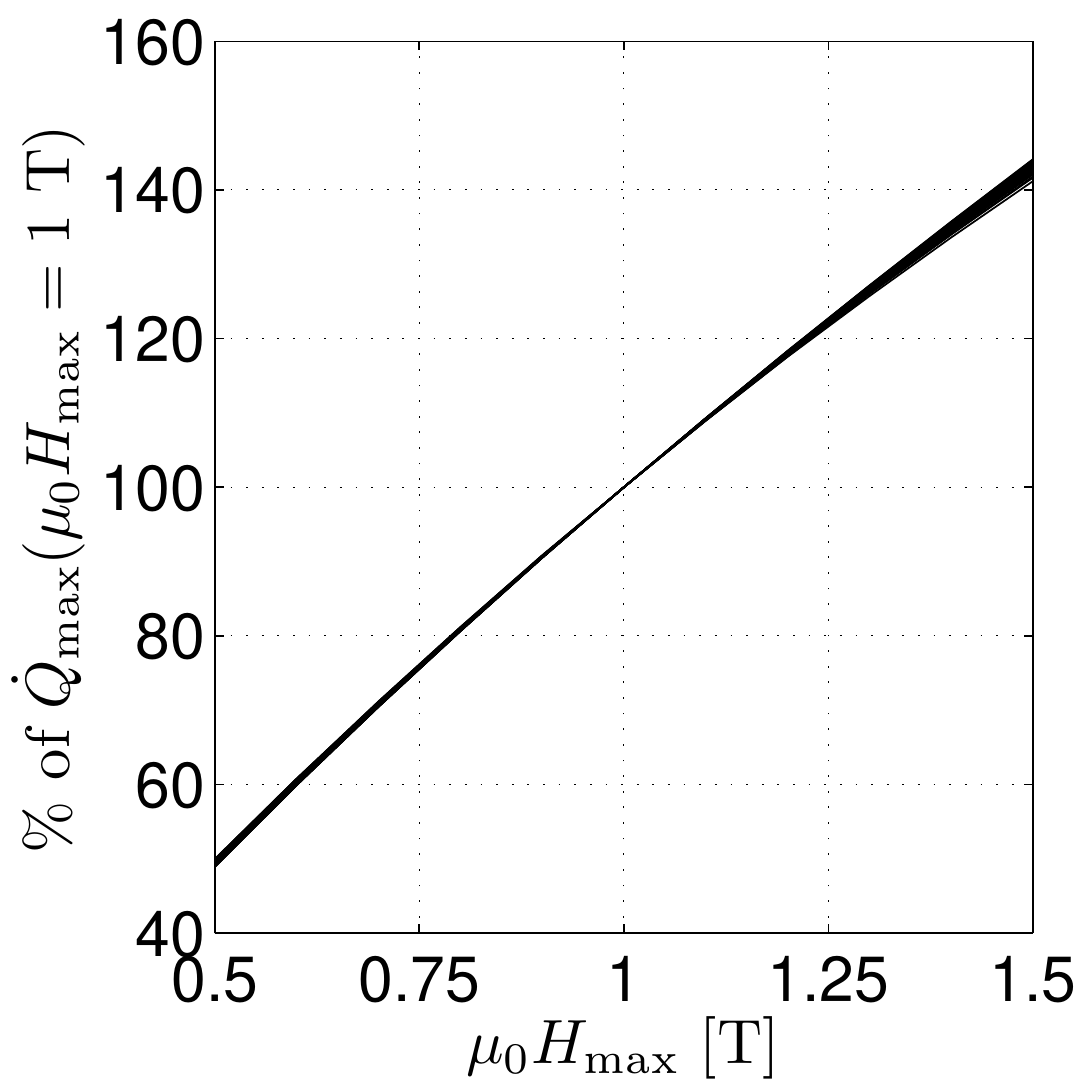}}\hspace{0.4cm}
\subfigure[Packed bed.]{\includegraphics[width=0.47\textwidth]{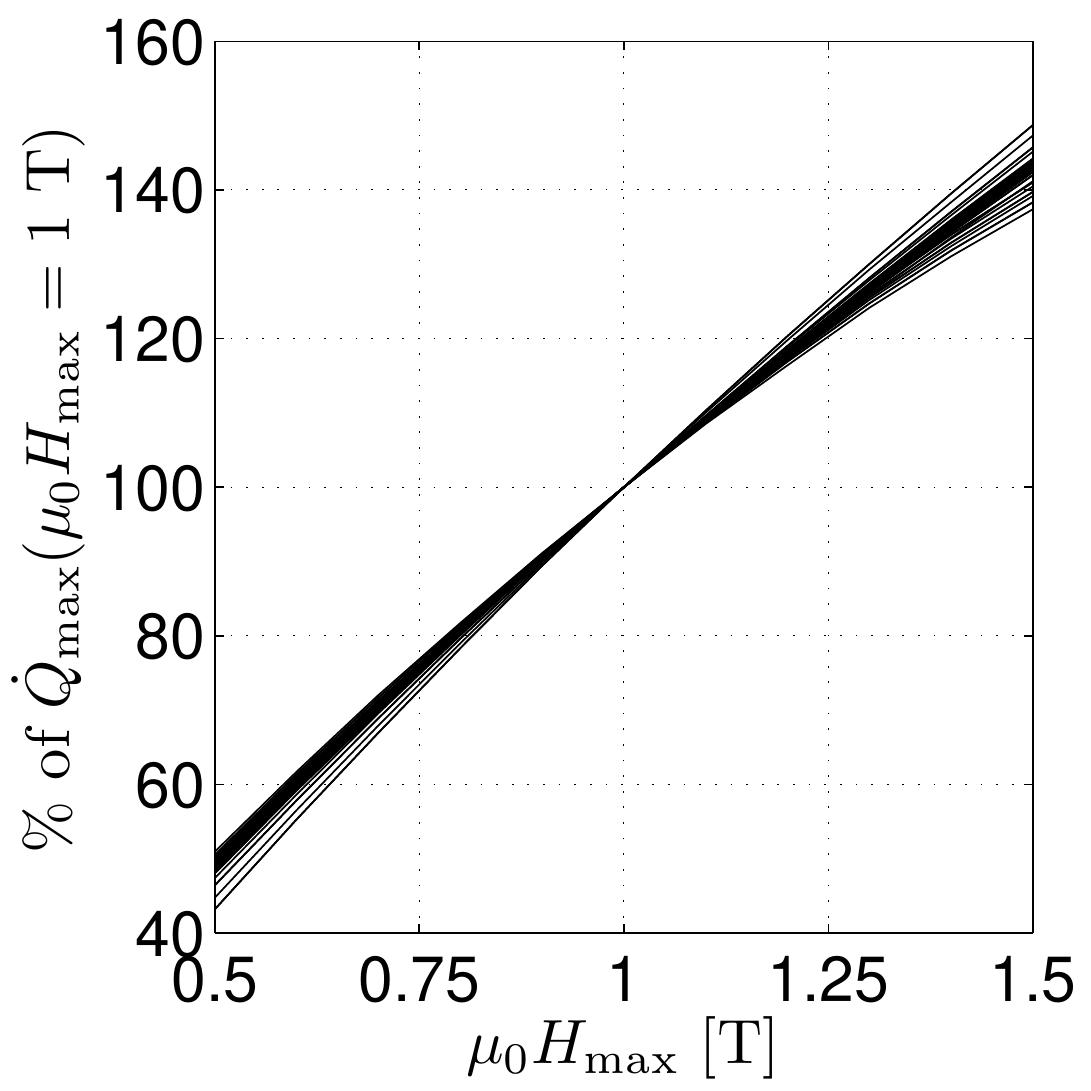}}
\end{center}
\caption{The maximum cooling capacity, $\dot{Q}_\n{max}$, as a function of $\mu_0H_\mathrm{max}$ for all parameter sets.}
\label{Fig.Field_Q_dot}
\end{figure*}

\begin{figure*}[!t]
\begin{center}
\subfigure[Parallel plates.]{\includegraphics[width=0.47\textwidth]{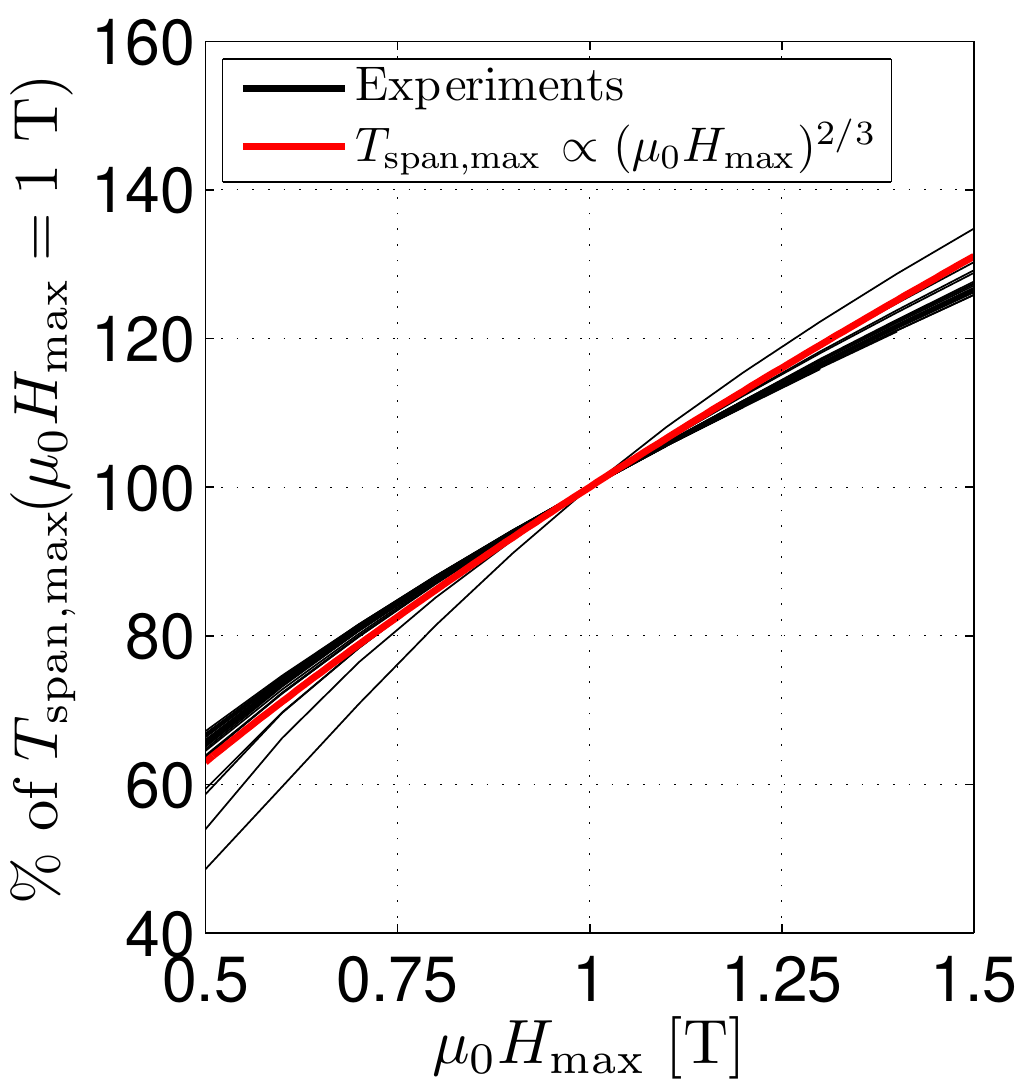}}\hspace{0.4cm}
\subfigure[Packed bed.]{\includegraphics[width=0.47\textwidth]{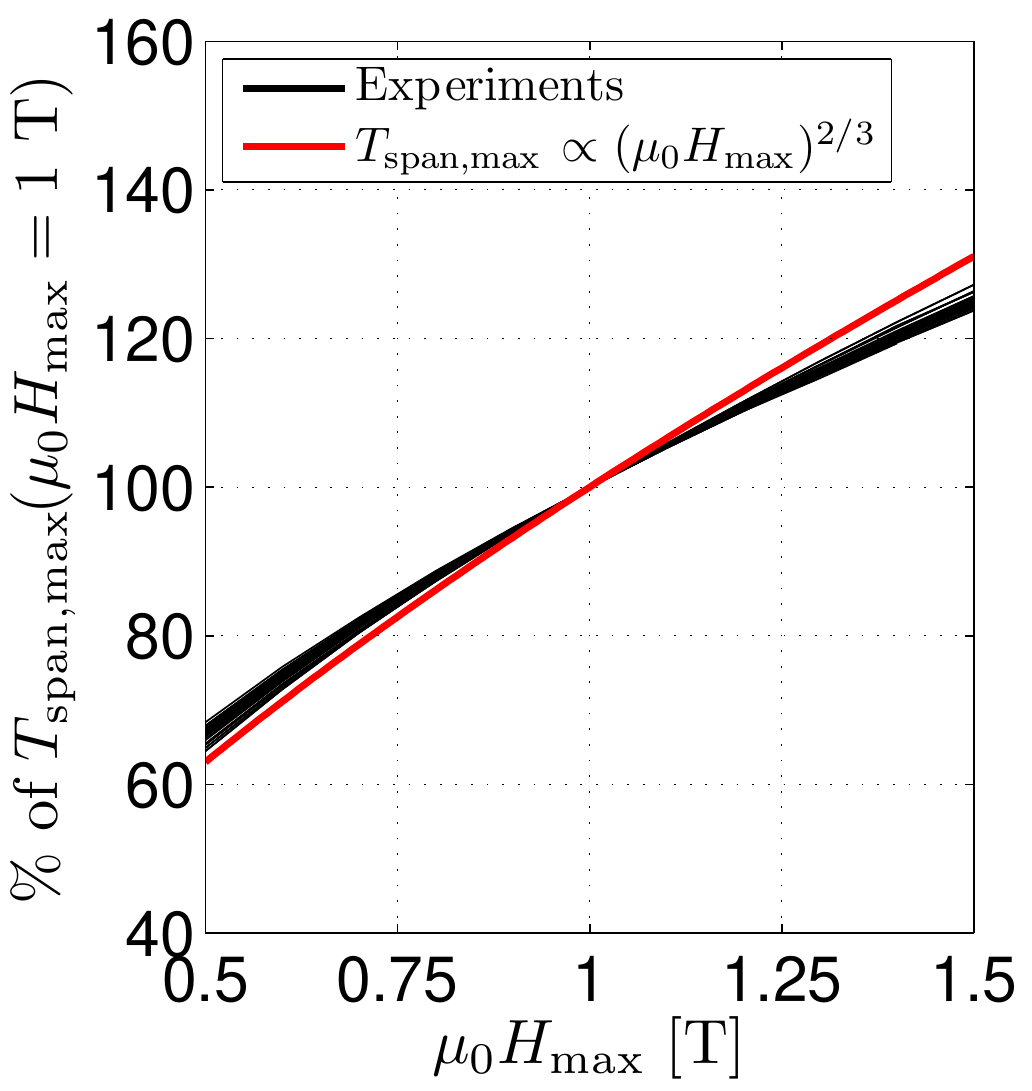}}
\end{center}
\caption{The no load temperature span, $T_\n{span,max}$, as a function of $\mu_{0}H_\mathrm{max}$ for all parameter sets. Also shown is $T_\n{span,max}\propto{}(\mu_{0}H_\n{max})^{2/3}$.}
\label{Fig.Field_T_span}
\end{figure*}

\section{Maximum value of the magnetic field}
Having determined the effect of synchronization and ramping of the magnetic field on the performance of the AMR, we now turn to the study of the effect of the maximum value of the magnetic field, $\mu_{0}H_\n{max}$.

To study this, the magnetic field profile is chosen such that $w_\n{top}=0.45$, $w_\n{total}=0.55$ and $x_0=0$, i.e. a synchronized magnetic field profile. The value of $\mu_0{}H_\n{max}$ was varied from 0.5 to 1.5 T in steps of 0.1 T. The temperature of the cold end of the regenerator was varied from 230 to 260 K in steps of 5 K and from 260 to 298 K in steps of 1 K in order to find $T_\n{span,max}$ and $\dot{Q}_\n{max}$. The larger temperature interval of the cold end temperature was considered in order to improve the estimate of $T_\n{span,max}$ produced by the high values of the magnetic field.

In Figs. \ref{Fig.Field_Q_dot} and \ref{Fig.Field_T_span} the maximum cooling capacity and temperature span are plotted as functions of $\mu_0H_\mathrm{max}$ for all process parameters. As can be seen from the figures $\dot{Q}_\n{max}$ scales almost identically for the different process parameters. In all cases $\dot{Q}_\n{max}$ has a stronger dependency on $\mu_0H_\n{max}$ than $T_\n{span,max}$. Thus increasing the magnetic field increases $\dot{Q}_\n{max}$ more than $T_\n{span,max}$, e.g. increasing $\mu_0H_\n{max}$ from 1 T to 1.5 T increases $\dot{Q}_\n{max}$ by $\sim{} 40\%$ but only $T_\n{span,max}$ by $\sim{} 25\%$.  The slope of increase for both  $T_\n{span,max}$ and $\dot{Q}_\n{max}$ with respect to $\mu_0H_\n{max}$ is below 1 for every parameter set modeled. This is due to the fact that, for mean field gadolinium, the adiabatic temperature change at the Curie temperature scales as a power law with an exponent of 2/3, i.e. $\Delta{}T_\n{ad}(T_\n{c}) = (\mu_0H)^{2/3}$ \citep{Oesterreicher_1984}. This is comparable with results from actual magnetocaloric materials \citep{Bjoerk_2010d}. This scaling, i.e. $\Delta{}T_\n{ad}(T_\n{c}) \propto (\mu_0H)^{2/3}$, is also shown. As can be seen $T_\n{span,max}$ scales with an exponent that is slightly less than $2/3$.

\begin{figure*}[!t]
\begin{center}
\subfigure[Parallel plates.]{\includegraphics[width=0.47\textwidth]{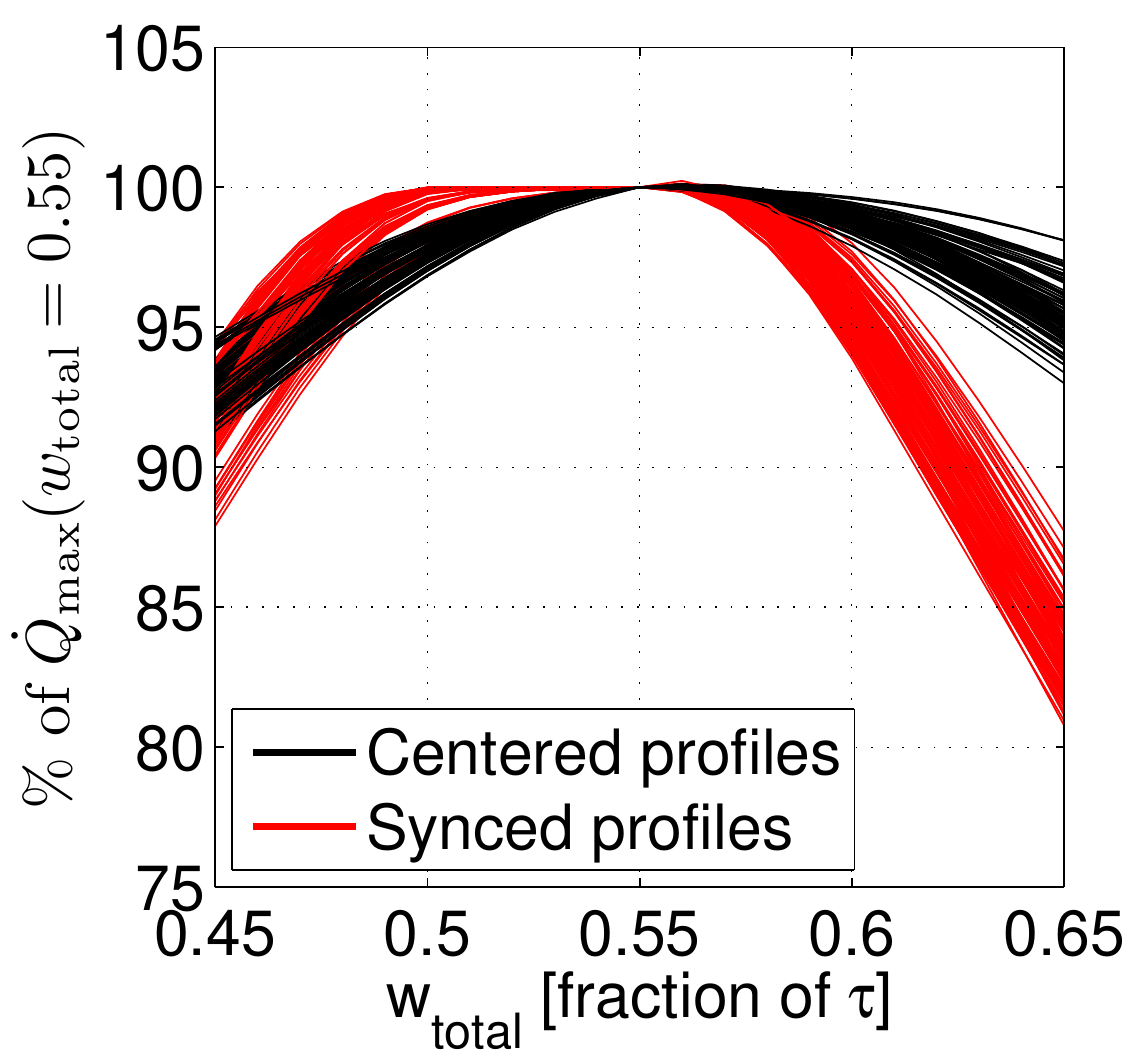}}\hspace{0.4cm}
\subfigure[Packed bed.]{\includegraphics[width=0.47\textwidth]{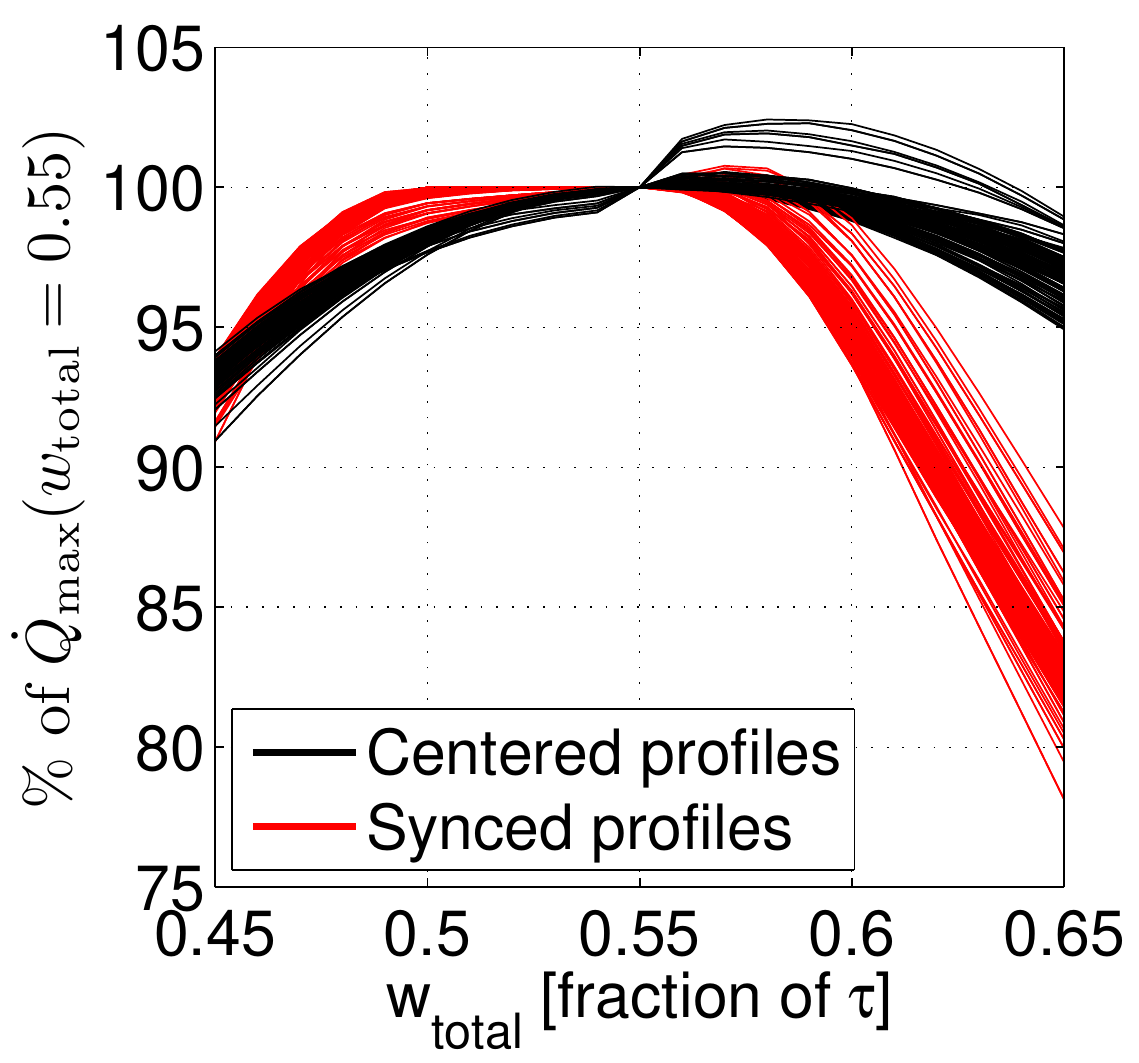}}
\end{center}
\caption{The maximum cooling capacity, $\dot{Q}_\n{max}$, as a function of $w_\n{total}$ for the parallel plate and packed bed cases. Both centered profiles and synced profiles are shown.}
\label{Fig.Shift_Q_dot_Length}
\end{figure*}

\begin{figure*}[!t]
\begin{center}
\subfigure[Parallel plates.]{\includegraphics[width=0.47\textwidth]{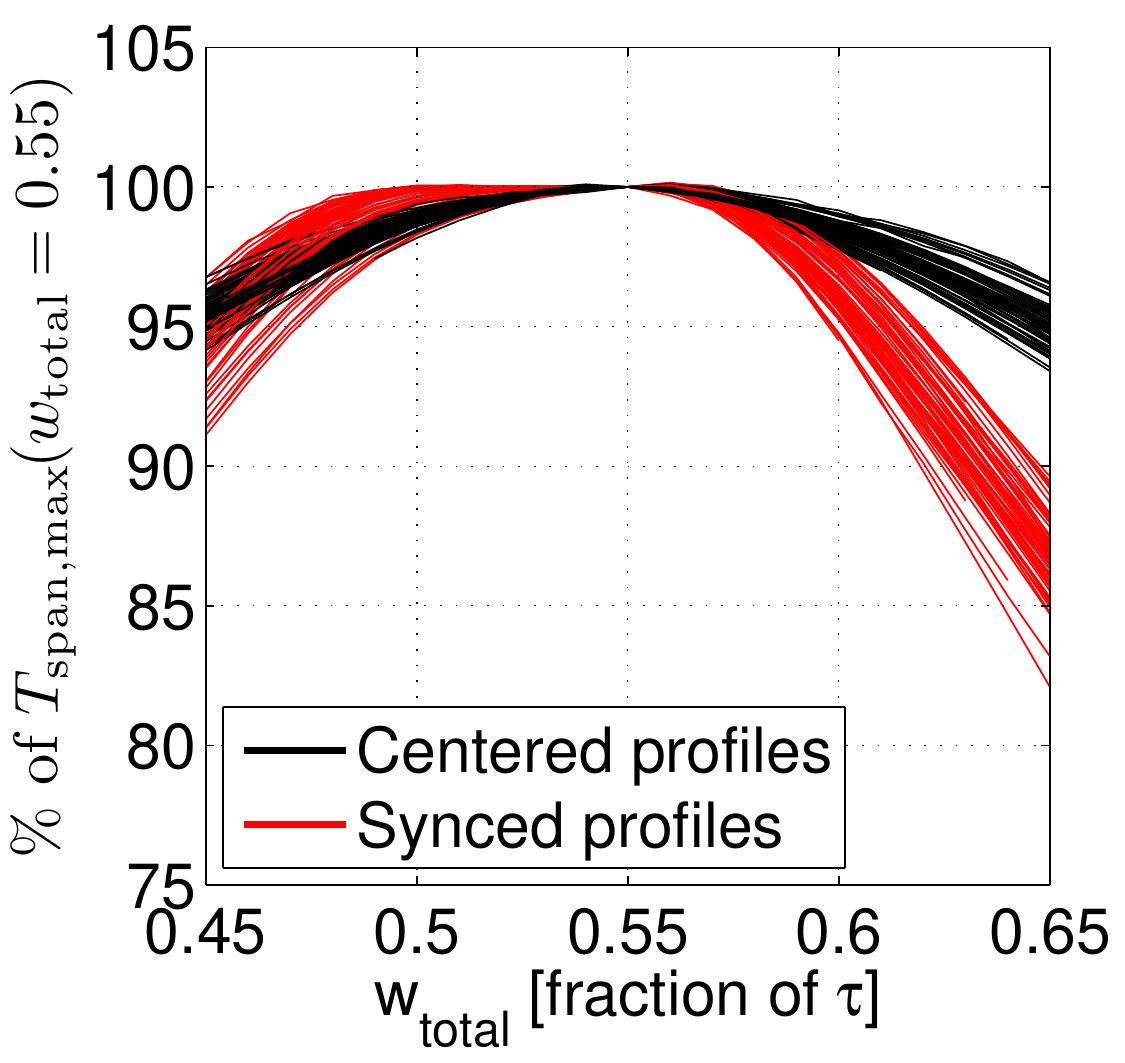}}\hspace{0.4cm}
\subfigure[Packed bed.]{\includegraphics[width=0.47\textwidth]{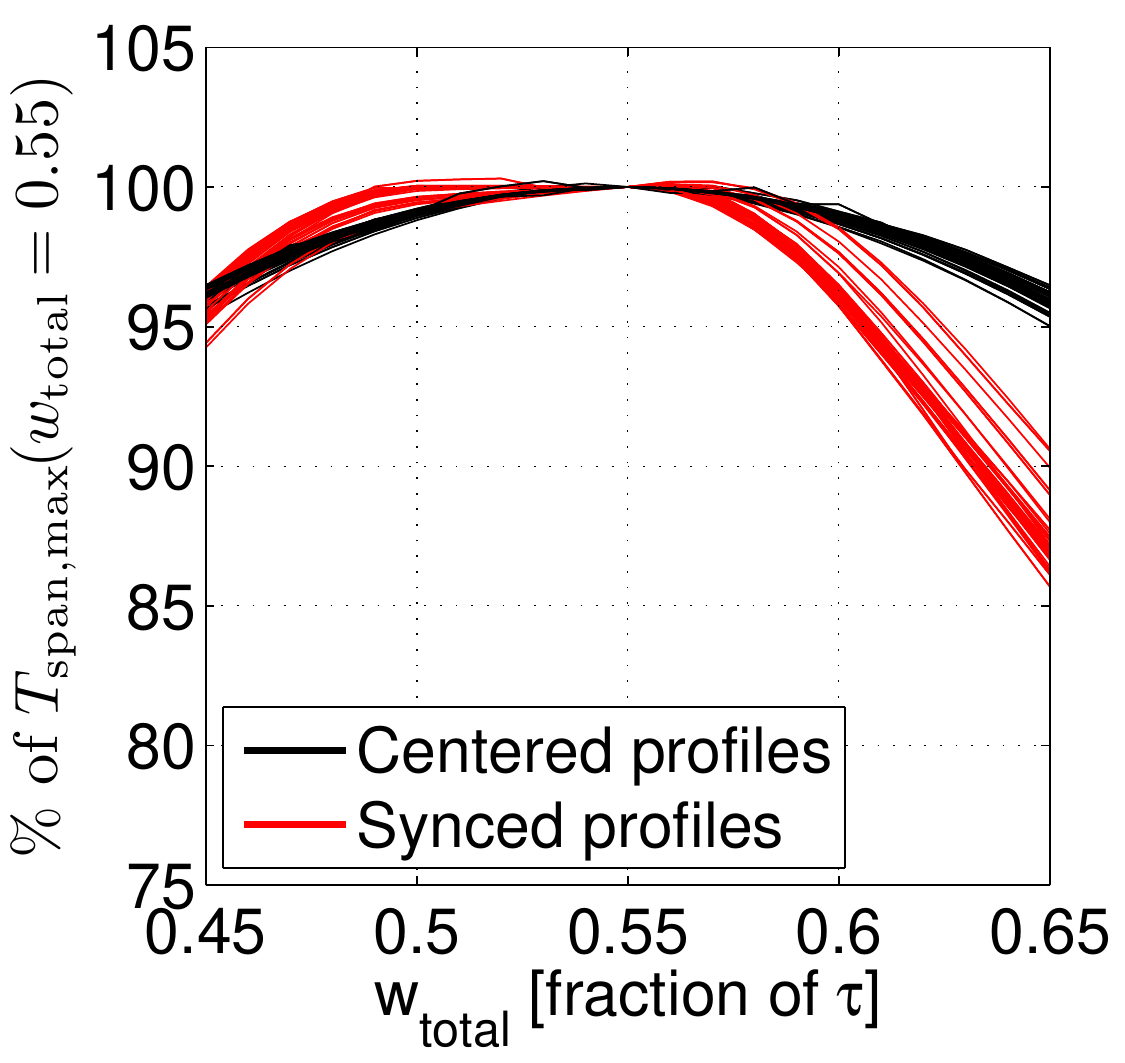}}
\end{center}
\caption{The no load temperature span, $T_\n{span,max}$, as a function of $w_\n{total}$ for the parallel plate and packed bed cases. Both centered profiles and synced profiles are shown.}
\label{Fig.Shift_T_span_Length}
\end{figure*}

\section{Width of the magnetic field}
It is also of importance to examine the behavior of the AMR cycle with respect to the width of the magnetic field profile, i.e. the fraction of the AMR cycle in which the regenerator is subjected to the high magnetic field or correspondingly how long the regenerator is in the low field region.  Here we consider a profile where the ramp time is kept constant at 5 \% of the total cycle time, i.e. $(w_\n{total}-w_\n{top})/2 = 0.05$. The $w_\n{top}$ parameter is varied from 0.35 to 0.55 in steps of 0.01, with the $w_\n{total}$ parameter given by the ramp time, i.e. $w_\n{total} = w_\n{top} + 0.1$.

\begin{figure*}[!t]
\begin{center}
\subfigure[Parallel plates.]{\includegraphics[width=0.47\textwidth]{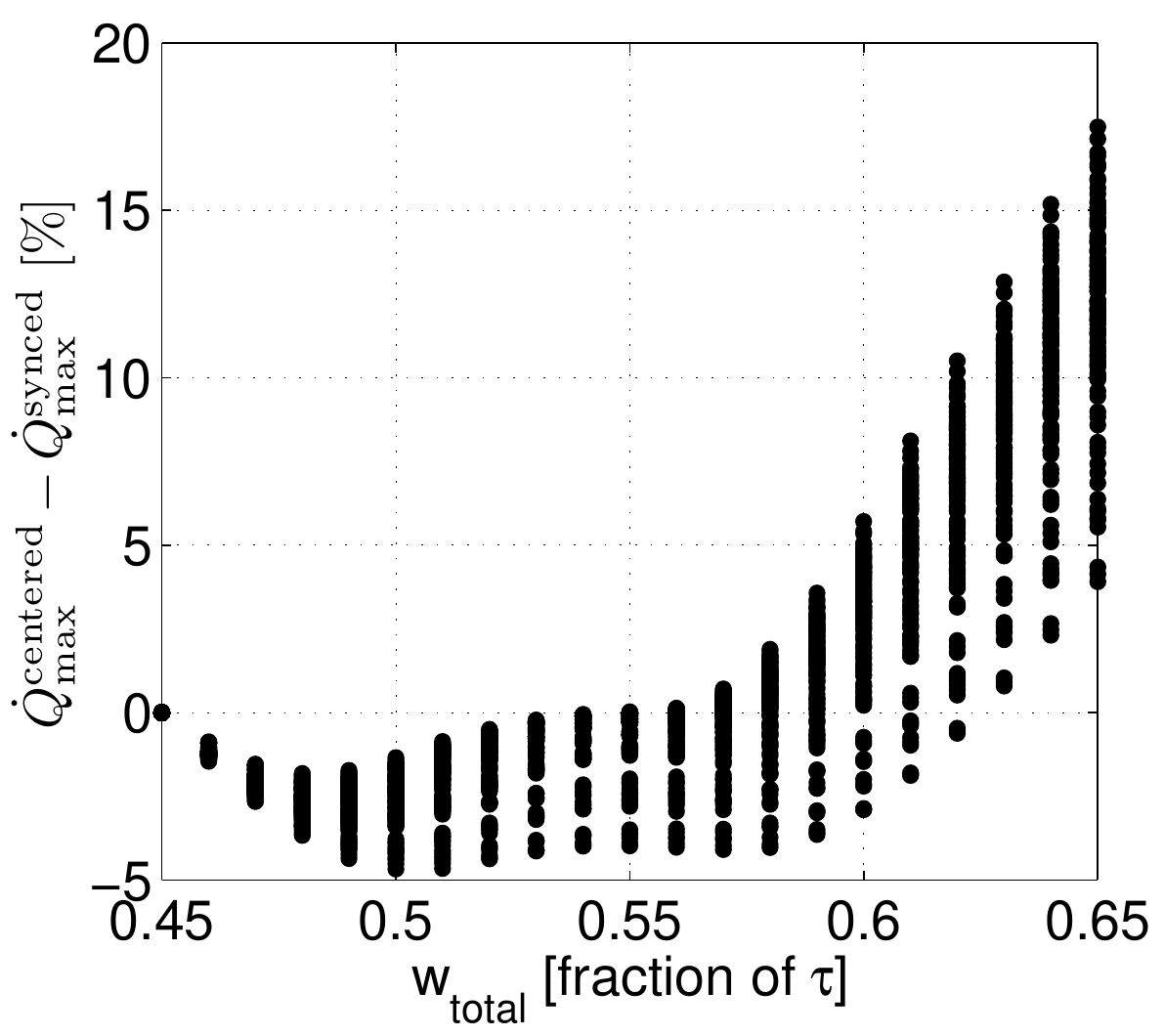}}\hspace{0.4cm}
\subfigure[Packed bed.]{\includegraphics[width=0.47\textwidth]{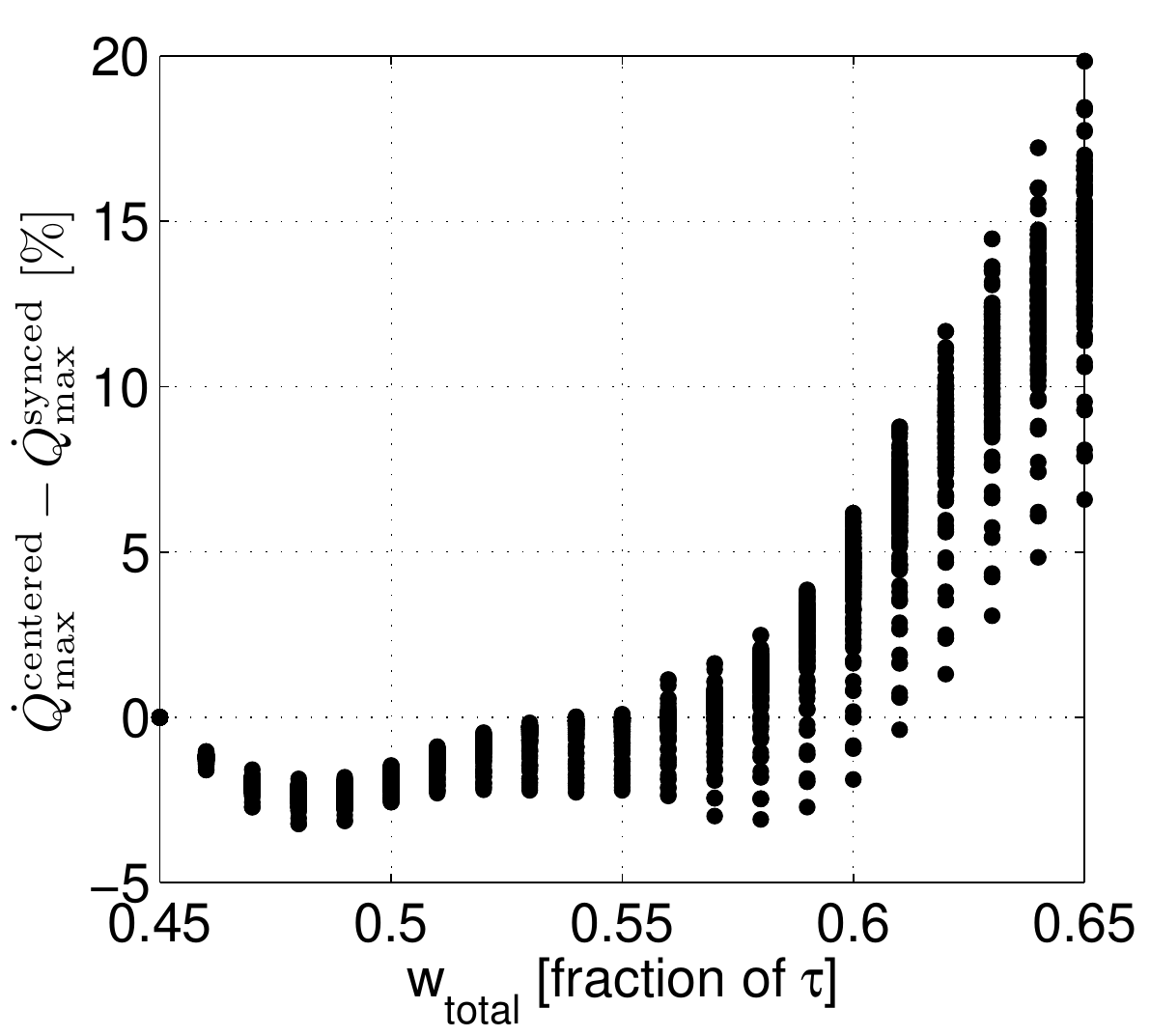}}
\end{center}
\caption{The difference in $Q_\n{max}$ between the synced and centered profiles as a function of $w_\n{total}$ for the parallel plate and packed bed cases.}
\label{Fig.Diff_synced_Q}
\end{figure*}

\begin{figure*}[!t]
\begin{center}
\subfigure[Parallel plates.]{\includegraphics[width=0.47\textwidth]{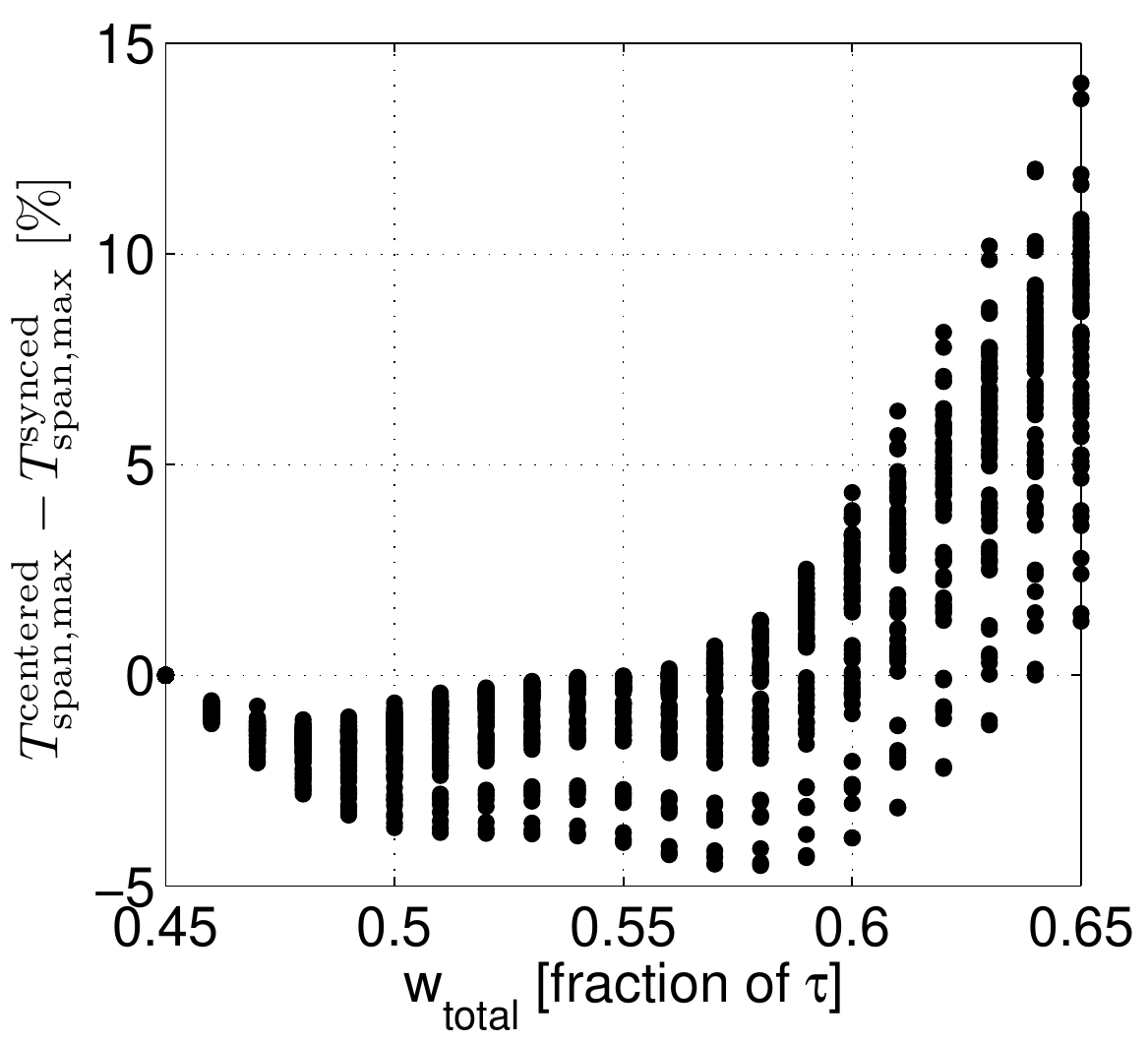}}\hspace{0.4cm}
\subfigure[Packed bed.]{\includegraphics[width=0.47\textwidth]{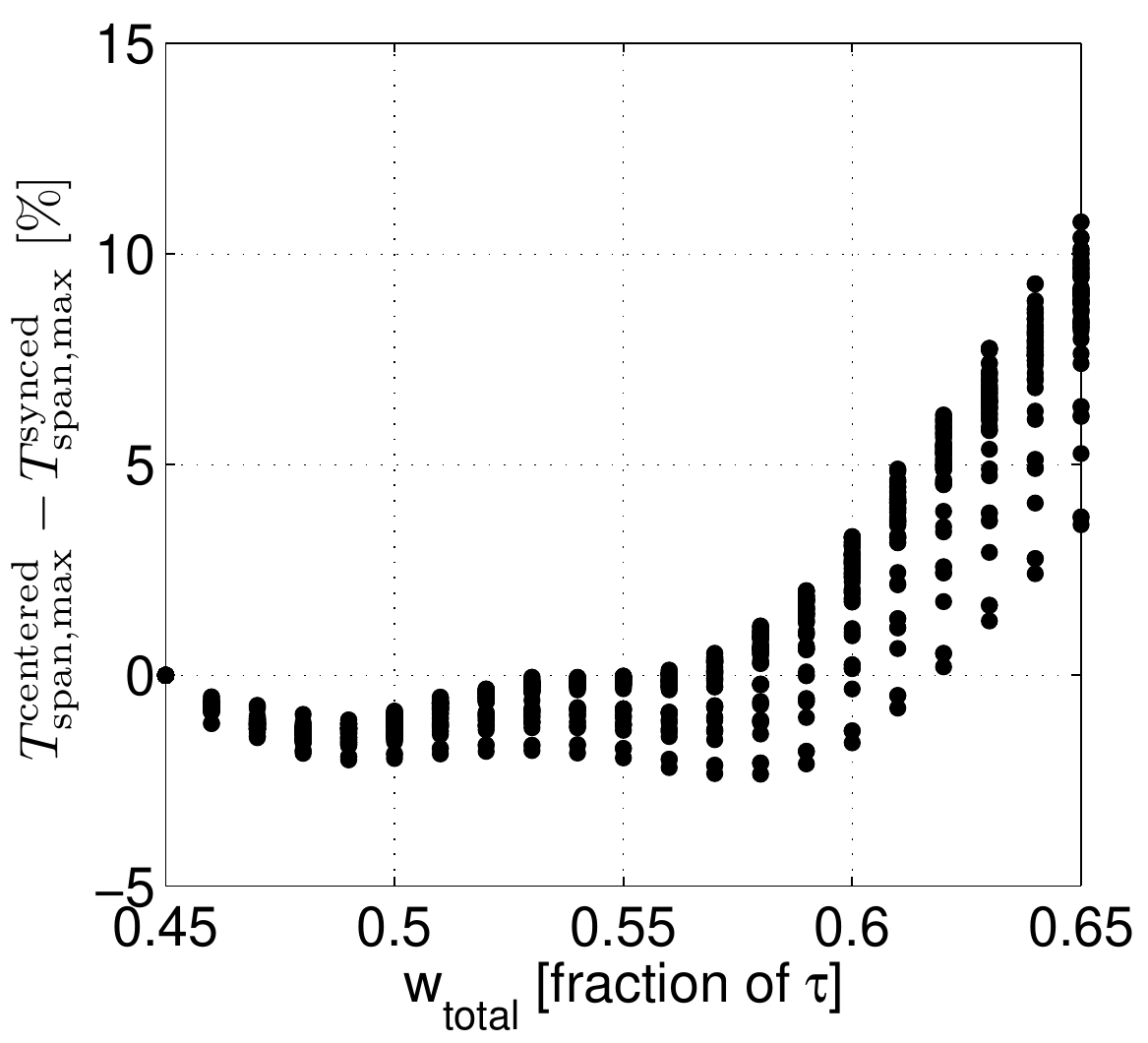}}
\end{center}
\caption{The difference in $T_\n{span,max}$ between the synced and centered profiles as a function of $w_\n{total}$ for the parallel plate and packed bed cases.}
\label{Fig.Diff_synced_T}
\end{figure*}

We consider two cases: one case where the magnetic field profile is centered on the flow profile, and one case where $x_0$ is changed so that the magnetic field profile begins to ramp down at the same time as the flow profile changes from $\dot{m}_\n{amp}$ to 0, i.e. at $t=\tau_1+\tau_2$. The first type of profile will be referred to as the centered profile, whereas the latter will be referred to as the synced profile.

\begin{table*}[!tb]
\begin{center}
\caption{The performance of the different tested profiles for a magnetic field profile characterized by the value in the ``Change'' column relative to the performance of a profile with the value given by ``Reference value'' column.}\label{Table.Performance_drop}
\begin{tabular}{l|c|c|ccccc}
   Case             & Reference & Change & \multicolumn{2}{c}{Performance: parallel}  & \multicolumn{2}{c}{Performance: packed} \\
                    & value & & \multicolumn{2}{c}{plate regenerator} & \multicolumn{2}{c}{bed regenerator}   \\
                    &           &              & $Q_\n{max}$ & $T_\n{span,max}$ & $Q_\n{max}$ &  $T_\n{span,max}$ \\ \hline
\multirow{2}{*}{Synchronization}     & \multirow{2}{*}{$x_0$ = 0} & $x_0$ = -0.1 & 60-75\%    & 60-80\%          &  60-75\%   & 70-80\% \\
     & & $x_0$ = \;\,0.1  & 80-95\%    & 80-95\%          &  85-90\%   & 90-95\% \\
&&& \\
\multirow{2}{*}{Ramp rate}           & \multirow{2}{*}{Rate = 0.05} & Rate = 0.005 & 96-100\%    & -   &  97-100\%   & -  \\
           &  & Rate = 0.1   & 96-98\%     & -   &  96-98\%    & -  \\
&&& \\
\multirow{2}{*}{Maximum field}    & \multirow{2}{*}{$\mu_0H_\mathrm{max}$ = 1 T} & $\mu_0H_\mathrm{max}$ = 0.5 T & 50\%     & 60-70\%   &  40-50\%   & 60-70\%  \\
    &  & $\mu_0H_\mathrm{max}$ = 1.5 T & 140\%  & 125\%   &  130-150\%   & 120-130\% \\
&&& \\
\multirow{2}{*}{Width (centered)}    & \multirow{2}{*}{$w_\mathrm{total}$ = 0.55} & $w_\mathrm{total}$ = 0.45 & 90-95\%     & 95-100\%   &  90-95\%   & 95-100\%  \\
    &  & $w_\mathrm{total}$ = 0.65 & 90-100\%     & 90-100\%   &  95-100\%   & 95-100\%  \\
&&& \\
\multirow{2}{*}{Width (synced)}    & \multirow{2}{*}{$w_\mathrm{total}$ = 0.55} & $w_\mathrm{total}$ = 0.45 & 85-95\%     & 90-95\%   &  90-95\%   & 95-100\%  \\
    &  & $w_\mathrm{total}$ = 0.65 & 80-90\%     & 85-90\%   &  80-90\%   & 85-90\%
\end{tabular}
\end{center}
\end{table*}

The fluid flow profile is kept fixed at the values introduced earlier, i.e. $\tau_1=\tau_3=0.1$ and $\tau_2=\tau_3=0.4$, respectively, in fractions of the total cycle time, $\tau$. Thus as $w_\n{top}$ is changed the width of the magnetic field profile will change from being ``shorter'' to ``longer'' than the fluid flow profile. This can be the case in e.g. a rotating AMR where the high field region can be smaller than the low field region or vice versa.

The modeling results are shown in Figs. \ref{Fig.Shift_Q_dot_Length} and \ref{Fig.Shift_T_span_Length}, which show $\dot{Q}_\n{max}$ and $T_\n{span,max}$ as a function of $w_\n{total}$. As can be seen from the figures, a more or less identical behavior is seen for all parameter sets. The optimum $w_\n{total}$ is approximately 0.55, and both $\dot{Q}_\n{max}$ and $T_\n{span,max}$ decrease as the value of $w_\n{total}$ moves further from 0.55. In general a large width is worse than a short width for the values considered here. The drop in performance for both $\dot{Q}_\n{max}$ and $T_\n{span,max}$ for $w_\n{total} > 0.55$ is greater for the synced profiles than for the centered profiles.

The difference between the centered and synced profiles for all parameter sets are shown in Fig. \ref{Fig.Diff_synced_Q} and \ref{Fig.Diff_synced_T}. For both $\dot{Q}_\n{max}$ and $T_\n{span,max}$ it is seen that if the magnetic field profile has a small value of $w_\n{total}$ it is better to sync the profile with the fluid flow profile, whereas for large values of $w_\n{total}$ the centered profile has the best performance.

\section{Discussion}
In Table \ref{Table.Performance_drop} the impact on performance is given for each of the magnetic field profile configurations tested.

For the all different magnetic field profiles tried it was seen that the influence of changing the magnetic field was more or less the same for the different regenerator geometries and operating parameters studied. This means that the design and optimization of the magnet can be done independently of the regenerator geometry. Thus, the optimum field strength and magnetized volume will be nearly the same for a packed sphere regenerator or parallel plate regenerator.

\section{Conclusion}
The influence of the magnetic field profile on the performance of an AMR was studied for different sets of process parameters. First, it was shown that when the hot and cold reservoir temperature are adequately far from the Curie temperature of the magnetocaloric material, the cooling curve is almost linear.  As the reservoir temperatures near the Curie temperature, the cooling curve flattens. Next, it was shown that a magnetic field profile that is 10\% of the total cycle time out of sync with the flow profile will lead to a drop in both the maximum temperature span and the maximum cooling capacity of 20-40\% for both parallel plate and packed bed regenerators. Also the maximum cooling capacity was shown to depend very weakly on the ramp rate of the magnetic field, whereas a drop in maximum temperature span and maximum cooling capacity of 5-20\% was seen when the temporal width of the magnetic field curve was changed by 10\%. It was shown that an increase of the magnetic field from 1 T to 1.5 T increased the maximum cooling capacity by 40\% but the maximum temperature span by only 25\%.  The relative change in performance caused by changing the magnetic field was found to be the same for the different regenerator geometries and operating conditions tried, which means that the magnet can be designed and optimized independently of the regenerator geometry. Finally, a magnetic field width of 0.55 was found to be optimum.

\section*{Acknowledgements}
The authors would like to thank Dr. C. R. H. Bahl for useful discussions. The authors would like to acknowledge the support of the Programme Commission on Energy and Environment (EnMi) (Contract No. 2104-06-0032) which is part of the Danish Council for Strategic Research.

\newpage
\bibliographystyle{elsarticle-harv}

\begin{thebibliography}{28}
\expandafter\ifx\csname natexlab\endcsname\relax\def\natexlab#1{#1}\fi
\expandafter\ifx\csname url\endcsname\relax
  \def\url#1{\texttt{#1}}\fi
\expandafter\ifx\csname urlprefix\endcsname\relax\def\urlprefix{URL }\fi

\bibitem[{Allab et~al.(2005)Allab, Kedous-Lebouc, Fournier, and
  Yonnet}]{Allab_2005}
Allab, F., Kedous-Lebouc, A., Fournier, J., Yonnet, J., 2005. Numerical
  modeling for active magnetic regenerative refrigeration. IEEE Transactions on
  Magnetics 41~(10), 3757--3759.

\bibitem[{Bahl et~al.(2008)Bahl, Petersen, Pryds, Smith, and
  Petersen}]{Bahl_2008}
Bahl, C., Petersen, T., Pryds, N., Smith, A., Petersen, T., 2008. A versatile
  magnetic refrigeration test device. Review of Scientific Instruments 79~(9),
  093906.

\bibitem[{Barclay(1982)}]{Barclay_1982}
Barclay, J., 1982. The theory of an active magnetic regenerativ refrigerator.
  NASA STI/Recon Technical Report N 83, 34087.

\bibitem[{Barclay(1988)}]{Barclay_1988}
Barclay, J.~A., 1988. Magnetic refrigeration: a review of a developing
  technology. Advances in Cryogenic Engineering 33, 719--731.

\bibitem[{Bj\o{}rk et~al.(2010{\natexlab{a}})Bj\o{}rk, Bahl, and
  Katter}]{Bjoerk_2010d}
Bj\o{}rk, R., Bahl, C., Katter, M., 2010{\natexlab{a}}. Magnetocaloric
  properties of {LaFe$_{13-x-y}$Co$_x$Si$_y$} and commercial grade {Gd}.
  Journal of Magnetism and Magnetic Materials 322~(24), 3882-3888.

\bibitem[{Bj\o{}rk et~al.(2010{\natexlab{b}})Bj\o{}rk, Bahl, Smith, and
  Pryds}]{Bjoerk_2010b}
Bj\o{}rk, R., Bahl, C. R.~H., Smith, A., Pryds, N., 2010{\natexlab{b}}. Review
  and comparison of magnet designs for magnetic refrigeration. International
  Journal of Refrigeration 33, 437-448.

\bibitem[{Dan'kov et~al.(1998)Dan'kov, Tishin, Pecharsky, and
  Gschneidner}]{Dankov_1998}
Dan'kov, S., Tishin, A., Pecharsky, V., Gschneidner, K.A., J., 1998. Magnetic
  phase transitions and the magnetothermal properties of gadolinium. Physical
  Review B (Condensed Matter) 57~(6), 3478--3490.

\bibitem[{Engelbrecht(2008)}]{Engelbrecht_2008}
Engelbrecht, K., 2008. A numerical model of an active magnetic regenerator
  refrigerator with experimental validation. Ph.D. thesis, University of
  Wisconsin - Madison.

\bibitem[{Engelbrecht et~al.(2005{\natexlab{a}})Engelbrecht, Nellis, and
  Klein}]{Engelbrecht_2005a}
Engelbrecht, K., Nellis, G., Klein, S., 2005{\natexlab{a}}. A numerical model
  of an active magnetic regenerator refrigeration system. Tech. Rep. Tech. Rep.
  ARTI-21CR/612-10075, University of Wisconsin-Madison.

\bibitem[{Engelbrecht et~al.(2005{\natexlab{b}})Engelbrecht, Nellis, Klein, and
  Boeder}]{Engelbrecht_2005b}
Engelbrecht, K., Nellis, G., Klein, S., Boeder, A., 2005{\natexlab{b}}.
  Modeling active magnetic regenerative refrigeration systems. Proceedings of
  the 1$^\mathrm{st}$ International Conference on Magnetic Refrigeration at
  Room Temperature, Montreux, Switzerland, 265--274.

\bibitem[{Engelbrecht et~al.(2007)Engelbrecht, Nellis, Sanford, and
  Zimm}]{Engelbrecht_2007c}
Engelbrecht, K., Nellis, G.~F., Sanford, A.~K., Zimm, C.~B., 2007. Recent
  developments in room temperature active magnetic regenerative refrigeration.
  HVAC\&R Research 13 (4), 525--542.

\bibitem[{Engelbrecht et~al.(2006)Engelbrecht, Nellis, and
  Klein}]{Engelbrecht_2006}
Engelbrecht, K.~L., Nellis, G.~F., Klein, S.~A., 2006. Predicting the
  performance of an active magnetic regenerator refrigerator used for space
  cooling and refrigeration. HVAC and R Research 12~(4), 1077--1095.

\bibitem[{Gschneidner and Pecharsky(2008)}]{Gschneidner_2008}
Gschneidner, K.A., J., Pecharsky, V., 2008. Thirty years of near room
  temperature magnetic cooling: Where we are today and future prospects.
  International Journal of Refrigeration 31~(6), 945--961.

\bibitem[{Hu and Xiao(1995)}]{Hu_1995}
Hu, J., Xiao, J., 1995. New method for analysis of active magnetic regenerator
  in magnetic refrigeration at room temperature. Cryogenics 35~(2), 101--104.

\bibitem[{Jacobs(2009)}]{Jacobs_2009}
Jacobs, S., 2009. Modeling and optimal design of a multilayer active magnetic
  refrigeration system. Proceedings of the 3$^\mathrm{rd}$ International
  Conference on Magnetic Refrigeration at Room Temperature, Des Moines, Iowa,
  USA, 267--274.

\bibitem[{Liu et~al.(2007)Liu, Sun, Wang, Zhao, and Shen}]{Liu_2007}
Liu, G., Sun, J., Wang, J., Zhao, T., Shen, B., 2007. A comparison study of the
  entropy changes in materials with and without short-range magnetic order.
  Journal of Physics Condensed Matter 19~(46), 466215.

\bibitem[{Morrish(1965)}]{Morrish_1965}
Morrish, A.~H., 1965. The Physical Priciples of Magnetism. John Wiley \& Sons,
  Inc.

\bibitem[{Nielsen et~al.(2009)Nielsen, Bahl, Smith, Bj\o{}rk, Pryds, and
  Hattel}]{Nielsen_2009a}
Nielsen, K., Bahl, C., Smith, A., Bj\o{}rk, R., Pryds, N., Hattel, J., 2009.
  Detailed numerical modeling of a linear parallel-plate active magnetic
  regenerator. International Journal of Refrigeration 32~(6), 1478--1486.

\bibitem[{Nielsen et~al.(2009)Nielsen, Bahl, Smith, Pryds, and
  Hattel}]{Nielsen_2010}
Nielsen, K., Bahl, C., Smith, A., Pryds, N., Hattel, J., 2010.
  A comprehensive parameter study of an active magnetic regenerator
  using a 2D numerical model. International Journal of Refrigeration 33~(4), 753--764.

\bibitem[{Oesterreicher and Parker(1984)}]{Oesterreicher_1984}
Oesterreicher, H., Parker, F., 1984. Magnetic cooling near curie temperatures
  above 300k. Journal of Applied Physics 55, 4334--4338.

\bibitem[{Okamura et~al.(2005)Okamura, Yamada, Hirano, and S.}]{Okamura_2005}
Okamura, T., Yamada, K., Hirano, N., S., N., 2005. Performance of a
  room-temperature rotary magnetic refrigerator. Proceedings of the
  1$^\mathrm{st}$ International Conference on Magnetic Refrigeration at Room
  Temperature, Montreux, Switzerland, 319--324.

\bibitem[{Pecharsky and Gschneidner~Jr(2006)}]{Pecharsky_2006}
Pecharsky, V.~K., Gschneidner~Jr, K.~A., 2006. Advanced magnetocaloric
  materials: What does the future hold? International Journal of Refrigeration
  29 (8), 1239--1249.

\bibitem[{Petersen et~al.(2008{\natexlab{a}})Petersen, Engelbrecht, Bahl,
  Pryds, Smith, Petersen, Elmegaard, and Engelbrecht}]{Petersen_2008b}
Petersen, T.~F., Engelbrecht, K., Bahl, C. R.~H., Pryds, N., Smith, A.,
  Petersen, T.~F., Elmegaard, B., Engelbrecht, K., 2008{\natexlab{a}}.
  Comparison between a 1d and a 2d numerical model of an active magnetic
  regenerative refrigerator. Journal of Physics D: Applied Physics 41~(10),
  105002.

\bibitem[{Petersen et~al.(2008{\natexlab{b}})Petersen, Pryds, Smith, Hattel,
  Schmidt, and H\o{}gaard~Knudsen}]{Petersen_2008a}
Petersen, T.~F., Pryds, N., Smith, A., Hattel, J., Schmidt, H.,
  H\o{}gaard~Knudsen, H.-J., 2008{\natexlab{b}}. Two-dimensional mathematical
  model of a reciprocating room-temperature active magnetic regenerator.
  International Journal of Refrigeration 31~(3), 432--443.

\bibitem[{Shir et~al.(2005)Shir, Mavriplis, Bennett, and Torre}]{Shir_2005b}
Shir, F., Mavriplis, C., Bennett, L.~H., Torre, E.~D., 2005. Analysis of room
  temperature magnetic regenerative refrigeration. International Journal of
  Refrigeration 28~(4), 616--627.

\bibitem[{Siddikov et~al.(2005)Siddikov, Wade, and Schultz}]{Siddikov_2005}
Siddikov, B., Wade, B., Schultz, D., 2005. Numerical simulation of the active
  magnetic regenerator. Computers \&amp; Mathematics with Applications
  49~(9-10), 1525--1538.

\bibitem[{Tura and Rowe(2007)}]{Tura_2007}
Tura, A., Rowe, A., 2007. Design and testing of a permanent
magnet magnetic refrigerator. Proceedings of the 2$^\mathrm{nd}$
International Conference of Magnetic Refrigeration at Room
Temperature, Portoroz, Slovenia, 363--370.


\bibitem[{Tura and Rowe(2009)}]{Tura_2009}
Tura, A., Rowe, A., 2009. Progress in the characterization and optimization of
  a permanent magnet magnetic refrigerator. Proceedings of the 3$^\mathrm{rd}$
  International Conference on Magnetic Refrigeration at Room Temperature, Des
  Moines, Iowa, USA, 387--392.

\bibitem[{Yu et~al.(2003)Yu, Gao, Zhang, Meng, and Chen}]{Yu_2003}
Yu, B., Gao, Q., Zhang, B., Meng, X., Chen, Z., 2003. Review on research of
  room temperature magnetic refrigeration. International Journal of
  Refrigeration 26~(6), 622--636.

\bibitem[{Zimm et~al.(2007)Zimm, Auringer, Boeder, Chell, Russek, and
  Sternberg}]{Zimm_2007}
Zimm, C., Auringer, J., Boeder, A., Chell, J., Russek, S., Sternberg, A., 2007.
  Design and initial performance of a magnetic refrigerator with a rotating
  permanent magnet. Proceedings of the 2$^\mathrm{nd}$ International Conference
  of Magnetic Refrigeration at Room Temperature, Portoroz, Slovenia, 341--347.

\end{thebibliography}

\end{document}